\documentclass[10pt,twocolumn,a4paper]{article}
\usepackage{graphicx}
\usepackage{fancyhdr}
\RequirePackage{amsmath,amsthm,amsfonts,amssymb}
\RequirePackage[colorlinks,linkcolor=blue,citecolor=blue,urlcolor=black]{hyperref}
\RequirePackage{graphicx}
\usepackage[square, sort&compress, numbers]{natbib}
\usepackage{algorithm}%
\usepackage{algorithmicx}%
\usepackage{algpseudocode}%
\usepackage{listings}%
\usepackage{enumerate}

\usepackage[T1]{fontenc}

\usepackage{booktabs}

\usepackage{caption}
\usepackage[margin=.6in]{geometry}

\newcommand{\scbf}[1]{\textbf{\textsc{#1}}}  
\newcommand{\E}{\mathbb{E}}

\newcommand{\var}{{\rm Var}}
\newcommand{\cov}{{\rm Cov}}

\newcommand{\pto}{\stackrel{P}{\to}}
\newcommand{\dto}{\stackrel{D}{\to}}

\newcommand{\iid}{\stackrel{{\footnotesize \rm i.i.d.}}{\sim}}

\newcommand{\limn}{\lim\limits_{n\to\infty}}

\newcommand{\methname}{AuxCov}
\newcommand{\regfname}{baseline}
\newcommand{\olsname}{\methname{}-OLS}
\newcommand{\splinesname}{\methname{}-Splines}
\newcommand{\glsname}{\methname{}-GLS}

\numberwithin{equation}{section}
\theoremstyle{plain}
\newtheorem{theorem}{Theorem}[section]

\newtheorem{corollary}{Corollary}[section]
\newtheorem{lemma}{Lemma}[section]
\newtheorem{assumptions}{Assumption}

\RequirePackage{times}

\newlength\FHoffset
\fancyheadoffset{\FHoffset}
\title{\Large\bf COVARIANCE MATRIX COMPLETION\\VIA AUXILIARY INFORMATION}

\author{Joseph Steneman and Giuseppe Vinci$^*$\\{\small \it Department of Applied and Computational Mathematics and Statistics}\\ {\small \it University of Notre Dame, Notre Dame, Indiana, USA}\\
  {\small \it$^*$Corresponding Author: gvinci@nd.edu}}
\date{}

\begin{document}

\twocolumn[
  \begin{@twocolumnfalse}
 
\maketitle

\abstract{Covariance matrix estimation is an important task in the analysis of multivariate data in disparate scientific fields. However, modern scientific data are often incomplete due to factors beyond the control of researchers, and traditional methods may only yield incomplete covariance matrix estimates. For example, it is impossible to obtain a complete sample covariance matrix if some variable pairs have no joint observations. We propose a novel approach, AuxCov, which exploits auxiliary variables to produce complete covariance matrix estimates from structurally incomplete data. In neuroscience, an example of an auxiliary variable is the distance between neurons, which is typically inversely related to the strength of the neuronal correlation. AuxCov estimates the relationship between the observed correlations and the auxiliary variables via regression, and uses it to predict the missing correlation estimates and to regularize the observed ones. We implement AuxCov using parametric and nonparametric regression methods, and propose procedures for tuning parameter selection and uncertainty quantification. We evaluate the performance of AuxCov through simulations and in the analysis of large-scale neuroscience data.}

~

\noindent
{\small {\sc Keywords:} asymptotic normality, Fisher transformation, measurement error, missing data, neuroscience, partial correlation.}

  \end{@twocolumnfalse}

  ~

  ~
]

\section{Introduction}\label{sec:intro}
Covariance matrices are useful to describe and quantify the dependence among many random variables. Covariance estimation is an important task in the analysis of multivariate data in disparate research fields, including neuroscience \citep{kohn2005stimulus,smith2013spatial,vinci2016separating,vinci2018adjusted,vinci2018adjustedcovariance}, genomics \citep{schafer2005shrinkage,hine2006determining,cai2013covariate,gan2022correlation}, astronomy \citep{taylor2013putting,grieb2016gaussian}, and finance \citep{ledoit2003improved,guo2017dynamic,deshmukh2020improved}, and has been a subject of research for several decades. Various approaches have been proposed for the estimation of large covariance matrices, including regularization techniques via constrained or penalized optimisation \citep{yuan2007model,friedman2008sparse,fan2009network,rothman2008sparse,bickel2008regularized, cai2010optimal,cai2011adaptive,bien2011sparse,chandrasekaran2012} and Bayesian methods \citep{wang2012bayesian, banerjee2015bayesian, mohammadi2015bayesian, wang2015scaling, vinci2018adjusted, vinci2018adjustedcovariance} often in the context of high-dimensional Gaussian graphical models, where the main goal is to estimate the inverse of the covariance matrix.

However, modern scientific data are often incomplete due to factors beyond the control of researchers, and data missingness may prohibit or impair the use of traditional covariance estimation methods. For example, many entries in a sample covariance matrix may be impossible to compute if the relative pairs of variables are not observed jointly. A traditional approach is to complete the data matrix first, assuming, for example, a low-rank structure \citep{candes2009exact, candes2012exact, candes2010matrix, cai2010singular}, and then proceed with the estimation of the covariance matrix based on the completed data. Alternatively, there exist methods that can complete a partially observed covariance matrix by assuming it to have a low-rank structure \cite{bishop2014deterministic}, or by imposing constraints on the inverse of the covariance matrix \citep{dempster1972covariance,grone1984positive,vinci2019graph}. 

We propose a new covariance matrix completion approach, \methname{}, which exploits \textit{auxiliary information}. Auxiliary variables are supplementary quantities that are available beyond the observed traditional data, and that may be informative about the unknown parameters of a statistical model. For example, in neuroscience, the strength of the dependence between neurons has been observed to decrease with inter-neuron distance and to increase with tuning curve similarity \citep{kohn2005stimulus,smith2013spatial,vinci2016separating}. 
\methname{} estimates the relationship between the observed correlations and the auxiliary variables through regression, and then uses it to predict the missing covariances and to regularize the observed ones. We propose various parametric and nonparametric approaches for the regression fitting, including ordinary least squares, generalized least squares, and splines. As an interesting auxiliary result, we demonstrate that the vector of Fisher-transformed observed sample correlations is asymptotically multivariate normal, and we derive a closed-form expression of its asymptotic covariance matrix. This result is important to model the dependence among the observed empirical correlations in the regression step of \methname{}, but it may be of interest in other contexts, including multiple hypothesis testing and the construction of confidence regions from incomplete data. 
We propose a cross-validation procedure for the selection of the \methname{} tuning parameters, and quantify the uncertainty of the \methname{} estimates via bootstrap.

Methods related to \methname{} were proposed by \cite{vinci2018adjusted, vinci2018adjustedcovariance}, who showed that inter-neuron distance and tuning curve similarity can be used to improve inverse covariance estimation, in the case of fully observed data. The settings we consider here are more challenging, because the data are not fully observed, and we exploit auxiliary variables to complete covariance matrix estimates. Another method related to ours was proposed in \cite{gan2022correlation} for the estimation of covariance matrices from single cell RNA-seq data. However, their approach applies ensemble learning to combine multiple auxiliary full covariance estimates into one, whereas auxiliary variables are not required to be covariances in our approach. \cite{zou2017covariance} and  \cite{fan2024covariance}  proposed to estimate the covariance matrix by assuming it to be a linear combination of several known matrices, and \cite{zheng2019hypothesis} proposed methods to test this modelling assumption. However, none of these methods deal with missing data, which is the focus of this paper. Finally, \methname{} requires no specific missingness pattern in the observed data, such as the minimum overlap between the observed data sets required in \cite{bishop2014deterministic}, and can be used to complete any incomplete covariance matrix estimate.

In Section~\ref{sec:assumptions}, we specify the setting of our covariance matrix completion problem. 
In Section~\ref{sec:estimation}, we present \methname{} with algorithms for its implementation and theoretical results, which are all proved in Appendix~\ref{app:proofs}. 
In Section~\ref{sec:simulations}, we present the results of an extensive simulation study to assess the performance of \methname{} and compare it with other existing methods. In Section~\ref{sec:data}, we apply \methname{} to the analysis of neuroscience data. Finally, in Section~\ref{sec:disc}, we discuss the importance of our proposed method.

\section{Problem statement}\label{sec:assumptions}

Let $X^{(1)}$, ..., $X^{(n)}$ be independent and identically distributed (i.i.d.) $p$-dimensional random vectors with mean vector $\mu=\E[X^{(1)}]\in\mathbb{R}^p$ and $p\times p$ positive definite covariance matrix $\Sigma=\cov(X^{(1)})\succ 0$, and let $V=\{1,\ldots,p\}$ be the set of variable indices. In this paper, we deal with the challenging situation in which none of these $n$ random vectors is fully observed. Specifically, let
\begin{equation}\label{eq:Vr}
V^{(r)}:=\{i\in V:X_i^{(r)}\text{ is observed}\}
\end{equation}
be the set of variables that are observed in the $r$-th sample,  
\begin{eqnarray}\label{eq:Nij}
\mathcal{N}_{ij} &:=& \{r\in\{1,...,n\}: \{i,j\}\subseteq V^{(r)}\}
\end{eqnarray}
be the sample indices where the variable pair $(i,j)$ was observed, and \vspace{-2mm}
\begin{eqnarray}\label{eq:nij}
n_{ij} &:=& |\mathcal{N}_{ij}|
\end{eqnarray}
be the joint sample size for the variable pair $(i,j)$. 
We will alternatively specify the observed data as a collection of $K$ data sets $\mathbf{X}^{(1)}, ..., \mathbf{X}^{(K)}$ with sample sizes $n_1,\ldots,n_K$ and $\sum_{k=1}^Kn_k=n$, about $K$ observed node subsets $V_1,...,V_K\subseteq V$. 
Thus, the set of jointly observed variable pairs is 
\begin{equation}\label{eq:O}
O~:=~\bigcup_{r=1}^nV^{(r)}\times V^{(r)}=\bigcup_{k=1}^KV_k\times V_k,
\end{equation}
and thereby $O^c$ is the set of variable pairs that have no joint observation. The relative size of $O^c$ is 
\begin{equation}\label{eq:eta}
    \eta ~:=~ \frac{|O^c|}{p^2}
\end{equation}
Assuming $\cup_{r=1}^n V^{(r)}=\cup_{k=1}^K V_k=V$, we have $\{(i,i); i\in V\}\subseteq O$, so each individual variable is observed. 

\begin{figure*}[t]
    \centering
\includegraphics[width=1\textwidth]{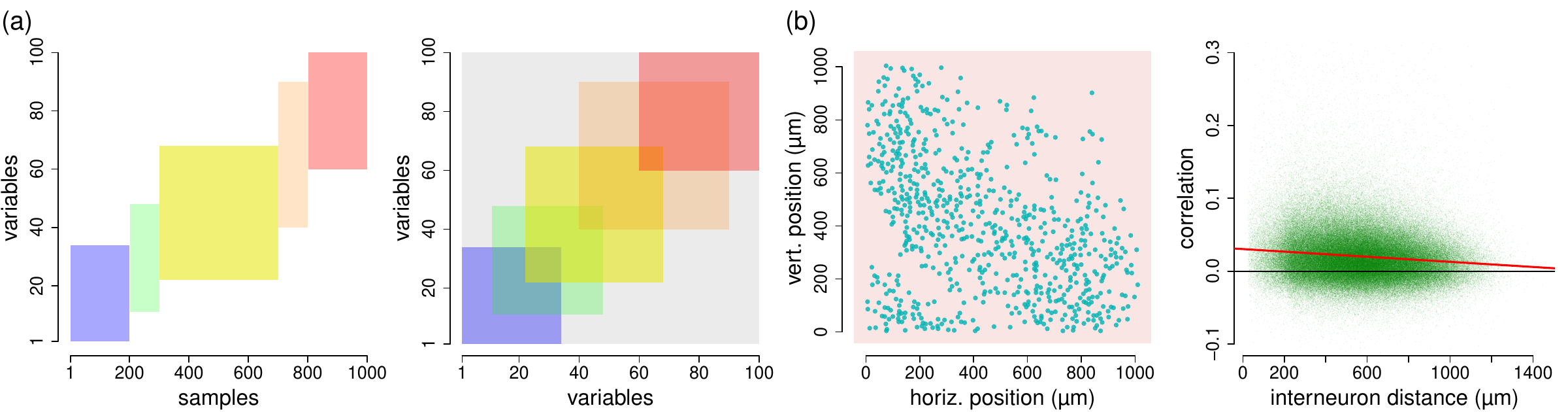}
    \caption{{\bf Incomplete data and auxiliary information}. 
    {\bf (a)} Observed subsets of 100 variables across 1000 samples and corresponding jointly observed variable pairs. Grey regions denote unobserved variable pairs ($\eta=0.4$). {\bf (b)} Positions of 725 neurons in mouse visual cortex and pairwise neuronal correlations plotted against inter-neuron distance. The average correlation decreases with inter-neuron distance, as highlighted by the fitted regression line (red).}
    \label{fig:motiv}
\end{figure*}

In Figure~\ref{fig:motiv}(a), we show an example observation pattern consisting of $K=5$ data sets about the variable subsets $V_1,...,V_5\subset \{1,...,100\}$ observed over a total of 1000 samples. A proportion $\eta=0.4$ of variable pairs have no joint observation (grey regions in the second panel).   

Since it is impossible to compute the empirical covariance of any variable pair $(i,j)\in O^c$, only an incomplete sample covariance matrix, $\hat\Sigma_O=\{\hat\Sigma_{ij}\}_{(i,j)\in O}$, may be computed, where
\begin{equation}\label{eq:obscov}
\hat\Sigma_{ij} ~:=~ \frac{1}{n_{ij}}\sum_{r\in \mathcal{N}_{ij}}(X_i^{(r)}-m_i)(X_j^{(r)}-m_j)
\end{equation}
and $m_k=\frac{1}{n_{kk}}\sum_{r\in \mathcal{N}_{kk}}X_k^{(r)}$, or $m_k=\mathbb{E}[X_k]$ if known. Under mild conditions, this estimator is consistent:
\begin{theorem}[\sc Consistency of the observed sample covariance]\label{theo:obscovconsist}
Let $X^{(1)},...,X^{(n)}$ be i.i.d.~$p$-dimensional random vectors with covariance matrix $\Sigma$. If $\Sigma_{ii}<\infty,\forall i$, then the observed sample covariance $\hat\Sigma_{ij}$ in Equation~\eqref{eq:obscov} is a consistent estimator of $\Sigma_{ij}$, $\forall(i,j)\in O$. 
\end{theorem}
While various covariance matrix completion methods exist (Section~\ref{sec:intro}), in this paper we propose a new flexible approach that leverages \textit{auxiliary information}. Auxiliary variables are supplementary quantities that are available beyond the observed traditional data and may be informative about the parameters of interest. For example, in neuroscience, the strength of the dependence between neurons has been observed to decrease with inter-neuron distance and increase with tuning curve similarity  \citep{kohn2005stimulus,smith2013spatial,vinci2016separating}, and in \cite{vinci2018adjusted} and \cite{vinci2018adjustedcovariance} it was shown that these two auxiliary quantities can be used to improve covariance matrix estimation in the case of fully observed data. 
In Figure~\ref{fig:motiv}(b), we show the positions of 725 neurons in mouse visual cortex (see Section~\ref{sec:data}) and their pairwise neuronal correlations plotted against inter-neuron distance. We can see that the average correlation decreases with inter-neuron distance. Several other auxiliary variables may be available in neuroscience and other fields, and they may be useful for covariance matrix completion. 

How can we use the relationship between correlations and auxiliary variables to obtain a complete covariance estimate of $\Sigma$ even if the data are incomplete? In the next section, we propose \methname{}, a novel approach that combines observed covariance estimates and auxiliary variables to yield a complete and regularized covariance matrix estimate.

\section{\methname}\label{sec:estimation}
The goal of our proposed method \methname{} is twofold. Firstly, given an incomplete covariance matrix estimate $\hat\Sigma_O$, we wish to use auxiliary variables to predict the missing covariance estimate of $\Sigma_{O^c}$. Secondly, we also want to use the auxiliary information to improve the estimation of $\Sigma_O$. In Section~\ref{sec:auxcovASS}, we state our main assumptions. In Section~\ref{sec:auxcovALGO}, we present the \methname{} algorithm. 
In Section~\ref{sec:estimF}, we provide methods and theory on the estimation of the relationship between observed correlations and auxiliary variables, an important step of the \methname{} algorithm.  In Section~\ref{sec:tuningparameter}, we discuss approaches for the data-driven selection of the tuning parameters of \methname{}. Finally, in Section~\ref{sec:auxcovUNCERT}, we propose bootstrap procedures for the uncertainty quantification of the \methname{} estimator.

\subsection{Assumptions}\label{sec:auxcovASS}

\methname{} builds upon the following assumptions on the observation pattern and 
on the relationship between correlations and auxiliary variables: 
\begin{assumptions}[\bf A1]
Let $X_1,\ldots, X_n$ be i.i.d.~$p$-dimensional random vectors with $p\times p$ covariance matrix $\Sigma$. In the $r$-th sample we observe only the variables in the set $V^{(r)}$, where $\cup_{r=1}^nV^{(r)}=V:=\{1,\ldots,p\}$. 
\end{assumptions}

\begin{assumptions}[\bf A2]
Let  $C = {\rm diag}(\Sigma)^{-1/2} \Sigma {\rm diag}(\Sigma)^{-1/2}$ be the correlation matrix. We assume
\begin{equation}\label{eq:auxrelat}
g(C_{ij}) ~=~ f(W_{ij})+\varepsilon_{ij}, ~~~1\le i<j\le p
\end{equation}
where $g:(-1,1)\to\mathbb{R}$ is an invertible function, $f$ is an unknown regression function, $W_{ij}\in\mathbb{R}^q$ is an observed $q$-dimensional auxiliary vector,  and $\varepsilon_{ij}$ is irreducible error. 
\end{assumptions}

Assumption (A1) characterizes the scenario of structural missingness described in Section 2, where multiple subsets of variables are observed independently, and many pairs of variables may lack joint observations. Assumption (A2) establishes a relationship between the covariance matrix and the auxiliary variables. Specifically, it posits a functional relationship between correlations and auxiliary variables, noting that correlations, as standardized versions of covariances, remain unaffected by variable scaling.

The \methname{} algorithm presented in the next section learns the regression function $f$ in Equation~\eqref{eq:auxrelat} from the data, by regressing the observed correlation estimates on the auxiliary variables. The function $g$ in Equation~\eqref{eq:auxrelat} maps correlations onto the real line, allowing for a more flexible and unconstrained modelling of the regression function $f$. In this paper, we take $g$ to be the Fisher transformation
\begin{equation}\label{eq:fishertransf}
    g(\varrho)=\tfrac{1}{2}\log\left(\tfrac{1+\varrho}{1-\varrho}\right),
\end{equation}
where $\log$ denotes the natural logarithm. The Fisher transformation is commonly applied to correlation coefficients to facilitate the implementation of statistical analyses such as hypothesis testing and confidence intervals.

\subsection{The \methname{} algorithm}\label{sec:auxcovALGO}

\begin{algorithm}[t]\small
\scbf{Input}:  Incomplete covariance matrix estimate $\hat{\Sigma}_{O}=\{\hat\Sigma_{ij}\}_{(i,j)\in O}$; $q$-dimensional auxiliary vectors $\{W_{ij}\}_{1\le i<j\le p}$; tuning parameter $\alpha \in [0,1]$; regression model $f:\mathbb{R}^q\to\mathbb{R}$; invertible function $g:(-1,1)\to\mathbb{R}$.\;
~

 \begin{enumerate}
    \item Compute the \textit{observed correlation estimates} $\hat{C}_{O}=\{\hat C_{ij}\}_{(i,j)\in O}$, where 
    \begin{equation}\label{eq:corrobs}
    \hat C_{ij}=\hat\Sigma_{ij}(\hat\Sigma_{ii}\hat\Sigma_{jj})^{-\frac{1}{2}}
    \end{equation}
    \item Obtain estimate $\hat f$ of the {\it baseline regression function}  $ f$ by regressing $g(\hat C_{ij})$ on $W_{ij}, (i,j)\in U$ (Equation~\eqref{eq:U}).
    \item Obtain the $p\times p$  \textit{auxiliary baseline matrix} $\bar{C}$, where
\begin{equation}\label{eq:corrbaseline}
\bar C_{ij} = \left\{
\begin{array}{cl}
  g^{-1}(\hat{f}(W_{ij})),   &  \text{if }i\neq j \\
  1,   &  \text{if }i=j
\end{array}
\right.
\end{equation}
    \item Obtain the $p\times p$ \textit{completed matrix} $\tilde{C}$, where 
        \begin{equation}
        \tilde C_{ij} = \left\{
        \begin{array}{cl}
          \hat C_{ij},   &  \text{if }(i,j)\in O \\
          \bar C_{ij},   &  \text{if }(i,j)\in O^c
        \end{array}
        \right.
        \end{equation}
        
    \item Apply positive definite correction (Algorithm~\ref{algo:posdiagadd}, Appendix~\ref{app:poscorrection}) on $\bar C$ and $\tilde C$, and then compute the $p\times p$ \textit{final correlation matrix estimate} 
        \begin{equation}\label{eq:CORalpha}
            \hat{C}(\alpha) = 
        \alpha \bar C + (1 - \alpha)\tilde C
        \end{equation}

\end{enumerate}
\scbf{Output}: Complete covariance matrix estimate 
\begin{equation}\label{eq:COValpha}
    \hat \Sigma(\alpha) = {\rm diag}(\hat\Sigma_O)^{\frac{1}{2}}\hat C(\alpha) {\rm diag}(\hat\Sigma_O)^{\frac{1}{2}}
\end{equation}
 \caption{\methname{}}\label{algo:algo1}
\end{algorithm}

\methname{} exploits Equation~\eqref{eq:auxrelat} to produce a complete estimate of the covariance matrix $\Sigma$ based on incomplete data and auxiliary variables. \methname{} is implemented in Algorithm~\ref{algo:algo1}, which takes as input an incomplete covariance matrix estimate $\hat\Sigma_O$ (e.g., the observed sample covariances in Equation~\eqref{eq:obscov}), auxiliary $q$-dimensional vectors $\{W_{ij}\}_{1\le i<j\le p}$, a tuning parameter $\alpha\in[0,1]$, a regression model $f:\mathbb{R}^q\to\mathbb{R}$, and an invertible function $g:(-1,1)\to\mathbb{R}$ (e.g., the Fisher transformation in Equation~\eqref{eq:fishertransf}). 
In step 1, we obtain the correlation estimates $\hat C_{ij}=\hat\Sigma_{ij}(\hat\Sigma_{ii}\hat\Sigma_{jj})^{-1/2}$, for all $(i,j)\in O$. 
In step 2, we estimate the \regfname{} regression function $f$ (Equation~\eqref{eq:auxrelat}) based on the observed pairs $\{(\hat C_{ij},W_{ij})\}_{(i,j)\in U}$, where 
\begin{equation}\label{eq:U}
    U:=\{(i,j)\in O: i<j\}
\end{equation}
is the upper diagonal entry set contained in $O$. We provide more details on this regression step in Section~\ref{sec:estimF}. 
In step 3, we obtain the $p\times p$ \textit{auxiliary baseline matrix} $\bar{C}$ by computing $\bar{C}_{ij} = g^{-1}(\hat{f}(W_{ij}))$ for $i \ne j$ and $\bar{C}_{ij} = 1$ for $i = j$. 
In step 4, we obtain a completed $p\times p$ correlation matrix $\tilde{C}$ as the union of $\hat C_O$ and $\bar C_{O^c}$. 
In step 5, we obtain the final correlation matrix estimate $\hat C(\alpha)$ as a convex combination of $\tilde C$ and $\bar C$, after positive definite correction if necessary (Algorithm~\ref{algo:posdiagadd}, Appendix~\ref{app:poscorrection}). The weights of the convex combination are governed by the tuning parameter $\alpha\in[0,1]$, which can be seen as a regularization parameter that, while not affecting the values of the predicted entries $\hat C_{O^c}$,  determines the shrinkage of the observed correlation estimates towards the auxiliary baseline values. The tuning parameter $\alpha$ can be selected via cross-validation, as described in Section~\ref{sec:tuningparameter}. 
Finally, the output of the algorithm is the final  estimated covariance matrix $\hat\Sigma(\alpha)$ obtained by rescaling $\hat C(\alpha)$ using the original observed variances $\hat\Sigma_{11},\ldots,\hat\Sigma_{pp}$. 

\begin{figure*}[ht]
    \centering
    \includegraphics[width=1\textwidth]{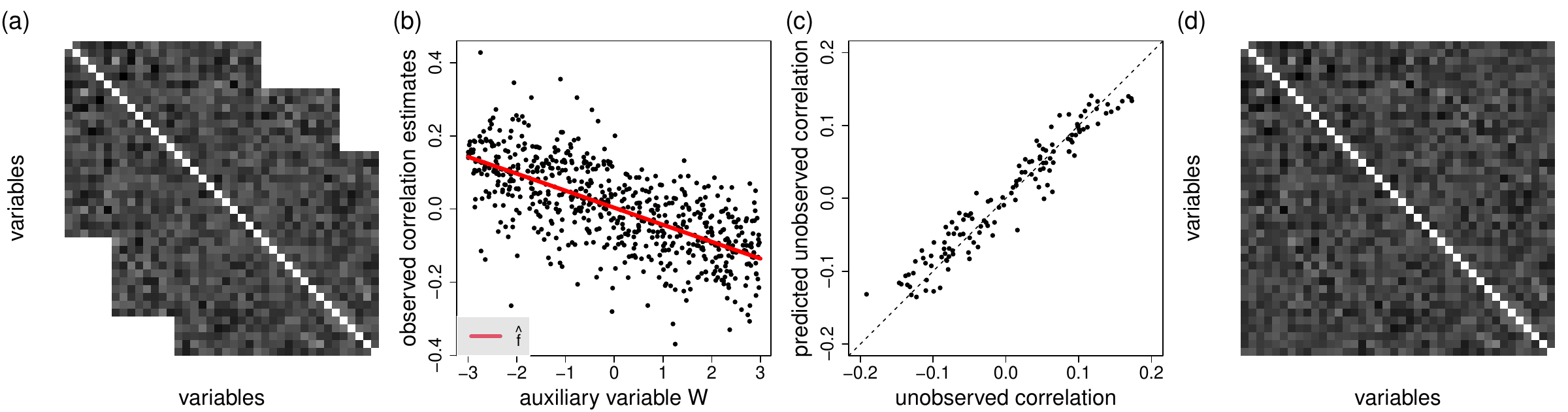}
    \caption{{\bf Illustration of the \methname{}  Algorithm~\ref{algo:algo1}}. 
    {\bf (a)} Incomplete covariance matrix estimate $\hat\Sigma_O$. 
    {\bf (b)} Regression of the available correlation estimates on the auxiliary variable $W$. 
    {\bf (c)} Predicted correlations $\bar C_{ij}$ in $O^c$ approximate well the  true correlations. 
    {\bf (d)} Complete \methname{} covariance matrix estimate $\hat\Sigma(\alpha)$.  
    }
    \label{fig:auxcovill}
\end{figure*}

In Figure~\ref{fig:auxcovill}, we illustrate the salient steps of Algorithm~\ref{algo:algo1}. In Figure~\ref{fig:auxcovill}(a), we display an incomplete covariance matrix estimate $\hat\Sigma_O$. In Figure~\ref{fig:auxcovill}(b), we plot the available correlation estimates $\hat C_{ij}$ (Equation~\eqref{eq:corrobs}) against an auxiliary variable $W_{ij}$, for $(i,j)\in U$, and the estimated baseline linear regression function $\hat f$. In Figure~\ref{fig:auxcovill}(c), we compare the true correlations $C_{ij}, (i,j)\in O^c$, with the predicted baseline correlations $\bar C_{ij}, (i,j)\in O^c$ (Equation~\eqref{eq:corrbaseline}). Finally, in Figure~\ref{fig:auxcovill}(d), we display the completed covariance matrix $\hat\Sigma(\alpha)$ (Equation~\eqref{eq:COValpha}), for some $\alpha\in[0,1]$.

\subsection{Baseline regression function estimation}\label{sec:estimF}
In this section, we propose various approaches to fitting the baseline regression function $f$ in Equation~\eqref{eq:auxrelat} in step 2 of the AuxCov Algorithm~\ref{algo:algo1}. The function $f$ specifies the relationship between the true correlation $C_{ij}$ and the auxiliary vector $W_{ij}$. However, in Algorithm~\ref{algo:algo1}, $C_{ij}$ is not observed, but rather we observe an estimate of it, $\hat C_{ij}$. Indeed, because of the sampling error of the correlation estimator, we have
\begin{equation}
    g(\hat C_{ij})=g(C_{ij})+\delta_{ij},
\end{equation}
where $\delta_{ij}$ may be regarded as a {\it measurement error}. Moreover, the correlation estimates, and thereby the measurement errors, may have quite different variances due to the different sample sizes in $O$, and are in general not independent. Indeed, for the observed empirical covariances, the following result holds:

\begin{lemma}\label{lemma:covsigmaO}
Let $X^{(1)},...,X^{(n)}$ be i.i.d.~$p$-dimensional random vectors with mean vector $\mu:=\E[X^{(1)}]$. Let $\hat\Sigma_{ij}$ be the observed empirical covariance between variables $(i,j)$ (Equation~\eqref{eq:obscov}) with $m_i:=\mu_i$, for $i=1,...,p$.  For any $(i,j),(k,l)\in O$, we have
\begin{equation}\label{eq:covcov}
\cov(\hat\Sigma_{ij},\hat\Sigma_{kl}) =   \frac{n_{ijkl}}{n_{ij} n_{kl}}   \cov\left( Z^{(1)}_i Z^{(1)}_j,Z^{(1)}_k Z^{(1)}_l\right)
\end{equation}
where $Z^{(r)}_i:=X^{(r)}_i-\mu_i$, $n_{ij}$ is the joint sample size for the variable pair $(i,j)$ (Equation~\eqref{eq:nij}), and $n_{ijkl}=|\mathcal{N}_{ij}\cap\mathcal{N}_{kl}|$ is the joint sample size for the variable quadruple $(i,j,k,l)$.
\end{lemma}
The proof of Lemma~\ref{lemma:covsigmaO} is in Appendix~\ref{app:proofs}. Regressing $g(\hat C_{ij})$ on $W_{ij}$ without taking into account the heteroskedastic and correlated measurement errors $(\delta_{ij})_{(i,j)\in U}$ can yield a distorted estimate of $f$. 
However, as shown via simulations in Section~\ref{sec:simulations}, fitting the baseline regression function $f$ while ignoring the measurement error (Section~\ref{sec:naive}) can still allow \methname{} to perform well compared to modelling the measurement error (Section~\ref{sec:fgls}). Moreover, accounting for the measurement error can be much more computationally expensive and yield more volatile estimates, especially for large $p$.

\subsubsection{Ignoring the measurement error}\label{sec:naive}
The simplest strategy is to ignore the measurement error, by fitting the regression 
\begin{equation}\label{eq:naivereg}
g(\hat C_{ij}) = f(W_{ij})+\varepsilon_{ij}, ~(i,j)\in U    
\end{equation}
where $U$ is defined in Equation~\eqref{eq:U}, and  treating $g(\hat C_{ij})$ as noiseless. In this article, we let $g$ be the Fisher transformation (Equation~\eqref{eq:fishertransf}) and fit $f$ linearly via ordinary least squares ({\bf \olsname{}}) or via splines ({\bf\splinesname{}}) with knots located at evenly spaced empirical quantiles of the predictor values and the number of knots selected via cross-validation (Algorithm~\ref{algo:cv}, Section~\ref{sec:tuningparameter}).

\subsubsection{Modelling the measurement error}\label{sec:fgls}
To account for measurement error in the estimation of the baseline regression function $f$, let us define the hierarchical model
\begin{equation}\label{eq:gls}
    \left\{
    \begin{array}{l}
    g(C_{ij}) = f(W_{ij})+\varepsilon_{ij}\\
    g(\hat C_{ij})= g(C_{ij})+\delta_{ij}  
    \end{array}
    \right., ~~ (i,j)\in U
\end{equation}
Assuming
\begin{eqnarray}
\mathcal{E}&:=&(\varepsilon_{ij})_{(i,j)\in U}~\sim~ N(0,\sigma_\varepsilon^2 I)\label{eq:Epsilon}\\
\Delta&:=&(\delta_{ij})_{(i,j)\in U}  ~\sim~ N(0,\Phi)\label{eq:Delta}
\end{eqnarray}
to be independent random vectors, and letting $\mathbf{W}$ be a $ |U|\times q$ matrix containing the values of $q$ auxiliary variables, we have $g(\hat C_U)=f(\mathbf{W})+\mathcal{E}+\Delta\sim N(f(\mathbf{W}),\sigma_\varepsilon^2 I+\Phi)$, where $g(\hat C_U)=(g(\hat C_{ij}))_{(i,j)\in U}$ and $f(\mathbf{W})=(f(W_{ij}))_{(i,j)\in U}$. If $\Phi$ is known, then an appropriately parameterized regression function $f_\beta$ and variance $\sigma_\varepsilon^2$ of the irreducible error can be estimated by maximum likelihood estimation (MLE). Algorithm~\ref{algo:optbaselinemeaserr} in Appendix~\ref{app:gls} implements this optimisation through gradient ascent for the case where $f_\beta$ is a linear regression function. The variant of \methname{} based on this approach is called {\bf \glsname{}}, because of the similarity to generalised least squares regression.

Of course, the covariance matrix $\Phi$ of the error vector $\Delta:=g(\hat C_U)-g(C_U)$ is generally unknown. However, the following theorem establishes that if $\hat C_U$ is the set of observed empirical correlations (Equation~\eqref{eq:obscov}) and $g(\varrho)$ is the Fisher transformation (Equation~\eqref{eq:fishertransf}), then, for large $n$, $\Delta$ is approximately Normally distributed in agreement with Equation~\eqref{eq:Delta}, and its asymptotic covariance matrix has a tractable closed-form expression:
\begin{theorem}[\sc Asymptotic Normality of Fisher-transformed observed sample correlations]\label{theo:asynormincomplete}
Let $X^{(1)},\dots,X^{(n)}$ be i.i.d.~$p$-dimensional random vectors with mean vector $\mu=\E[X^{(1)}]$, positive definite covariance matrix $\Sigma=\cov(X^{(1)})$, correlation matrix $C= {\rm diag}(\Sigma)^{-1/2} \Sigma {\rm diag}(\Sigma)^{-1/2}$, and $\E[(X_i^{(1)})^4]<\infty, \forall i$. Suppose that we observe only incomplete versions of these random vectors, specifically only variable subsets $V_1,...,V_K$ are observed over $n_1,...,n_K$ samples, respectively, where $n_k=\lceil n\cdot \pi_k\rfloor$, $\pi_k\in(0,1]$, and $\sum_{k=1}^K \pi_k=1$. Let $O$ and $U$ be defined as in Equations~\eqref{eq:O} and \eqref{eq:U}, let $\bar U:=\{(i,j)\in O: i\le j\}$, and let $g:(-1,1)\to\mathbb{R}$ be the Fisher transformation (Equation~\eqref{eq:fishertransf}). Let $\hat C_{ij}=\hat\Sigma_{ij}(\hat\Sigma_{ii}\hat\Sigma_{jj})^{-1/2}$ be the observed sample correlation based on the observed sample covariances $\hat\Sigma_O$ (Equation~\eqref{eq:obscov}). 
Then, as $n\to\infty$, 
\begin{equation}\label{eq:asympnormDelta}
\sqrt{n}\Delta~:=~\sqrt{n}(g(\hat C_{U}
)-g(C_{U}))\dto N(0,\Psi)
\end{equation}
where 
\begin{equation}\label{eq:PsiInc}
    \Psi =  F  JH J^{T}  F^{T}
\end{equation}  
and 
\begin{enumerate}[(i).]
    \item $H$ is a $|\bar U|\times |\bar U|$
    matrix with entry \begin{equation}\label{eq:Hentry}
    H_{(i,j),(k,l)}~=~c_{ijkl}\cdot\cov\big(Z^{(1)}_iZ^{(1)}_j,Z^{(1)}_kZ^{(1)}_l\big)
    \end{equation}
with $Z^{(1)}_i:=X^{(1)}_i-\mu_i$, 
\begin{eqnarray}\label{eq:cijkl}
c_{ijkl}&=&\frac{\sum_{t=1}^KI^{(t)}_{ij}I^{(t)}_{kl}\pi_t}{\left(\sum_{t=1}^KI^{(t)}_{ij}\pi_t\right)\left(\sum_{t=1}^KI^{(t)}_{kl}\pi_t\right)}~~~
    \end{eqnarray}
    and $I_{ij}^{(t)}:=\mathcal{I}\left(\{i, j\}\subseteq V_t\right) $, where $\mathcal{I}()$ is the indicator function, for $(i,j),(k,l)\in \bar U$; 
    
    \item 
    $ J $ is a $|U|\times |\bar U|$ matrix with entry $  J_{(i,j),(k,l)} = \frac{\partial (\Sigma_{ij}/\sqrt{\Sigma_{ii}\Sigma_{jj}})}{\partial\Sigma_{kl}}$, for $(i,j)\in U$ and $(k,l)\in\bar U$;
    
    \item ${F}$ is a $|U|\times|U|$ 
    diagonal matrix with entry $F_{(i,j),(i,j)} = (1-C_{ij}^2)^{-1} $, for $(i,j)\in U$.
\end{enumerate}
\end{theorem}
Theorem~\ref{theo:asynormincomplete} implies that, for large $n$, the covariance matrix $\Phi$ of the error vector $\Delta$ (Equation~\eqref{eq:Delta}) can be approximated by $\Psi/n$ (Equation~\eqref{eq:PsiInc}). Theorem~\ref{theo:asynormincomplete} assumes that the sample sizes of the $K$ data sets increase proportionally with $n$ according to $n_k=\lceil \pi_k\cdot n\rfloor$, where $\pi_1,\ldots,\pi_K\in(0,1]$ are constants. The scalar $c_{ijkl}$ in Equation~\eqref{eq:cijkl} also depends on the constants $\pi_1,...,\pi_K$, and is equal to 1 if the data are complete. Theorem~\ref{theo:asynormincomplete}  extends the result of \cite{neudecker1990asymptotic}, who derived the asymptotic distribution of the sample correlation matrix for the case of complete data. Indeed, the proof of Theorem~\ref{theo:asynormincomplete} (Appendix~\ref{app:proofs}) exploits techniques analogous to the ones used in \cite{neudecker1990asymptotic}, such as the Central Limit Theorem, Slutsky's Theorem, and the Delta Method. However, our proof involves numerous additional steps necessary to deal with incomplete data and Fisher transformation. Theorem~\ref{theo:asynormincomplete} may also be of interest in other contexts, including multiple hypothesis testing and the construction of confidence regions from incomplete data. Theorem~\ref{theo:asynormincomplete} is further verified via simulation in Appendix~\ref{app:veriftheo}. 

Since, for $(i,j),(k,l)\in\bar U$, $c_{ijkl}=0$ whenever the variable quadruple $(i,j,k,l)$ has not been observed jointly, all nonzero entries of the matrix $H$ and all entries of $J$ and $F$ depend only on distribution parameters related to the observed data. The following corollary takes advantage of this fact to specify two consistent estimators of $\Psi$ based only on the observed (incomplete) data:

\begin{corollary}\label{coro:psihat}
Under the conditions of Theorem~\ref{theo:asynormincomplete}, the following statements hold:
\begin{enumerate}[(i).]
\item If $\hat\Psi$ is a consistent estimator of $\Psi$ (Equation~\eqref{eq:PsiInc}), then
\begin{equation}
\sqrt{n}\hat\Psi^{-1/2}(g(\hat C_U)-g(C_U))\dto N(0,I_{|U|})
\end{equation}
\item A consistent empirical estimator of $\Psi$ is 
\begin{equation}\label{eq:psiemphat}
\hat\Psi=\hat F\hat J \hat H \hat J^T\hat F^T,
\end{equation}
where
{ \begin{eqnarray}\label{eq:Hemp}
\hspace{-12mm}    \hat H_{(i,j),(k,l)}\hspace{-3mm}  &=& \hspace{-3mm} \left\{
    \begin{array}{cc}
    \hspace{-2mm} \hat c_{ijkl}\left(M_{ijkl}-M_{ij}M_{kl}\right) & \hspace{-2mm}\text{if } \hat c_{ijkl}>0\\
    0 & \hspace{-2mm}\text{otherwise},
    \end{array}
\right.~~~~~~~\\
    \hat c_{ijkl} &=&\frac{n\cdot n_{ijkl}}{n_{ij}\cdot n_{kl}},\\
    M_{ijkl} &=& \frac{1}{n_{ijkl}}\sum_{r\in\mathcal{N}_{ij}\cap\mathcal{N}_{kl}}  Z_i^{(r)}Z_j^{(r)}Z_k^{(r)}Z_l^{(r)},\\
    M_{ij} &=& \frac{1}{n_{ij}}\sum_{r\in\mathcal{N}_{ij}} Z_i^{(r)}Z_j^{(r)},\\
    Z_i &=& X_i^{(r)}-M_i,\\
    M_i &=& \frac{1}{n_{ii}}\sum_{r\in\mathcal{N}_{ii}}X_i^{(r)},
    \end{eqnarray}}
and $\hat J$ and $\hat F$ are obtained by plugging $\hat\Sigma_O$ (Equation~\eqref{eq:obscov}) in place of $\Sigma_O$ in the matrices $J$ and $F$ (Equation~\eqref{eq:PsiInc}), respectively.

\item If $X^{(1)},\dots,X^{(n)}\iid N(\mu,\Sigma)$, then a consistent estimator of $\Psi$ is 
\begin{equation}\label{eq:psigausshat}
\tilde\Psi=\hat F\hat J \tilde H\hat J^T\hat F^T,
\end{equation}
where
{\begin{equation}
   \hspace{-12mm}     \tilde H_{(i,j),(k,l)} = \left\{
    \begin{array}{cc}
    \hat c_{ijkl}(\hat\Sigma_{ik}\hat\Sigma_{jl} + \hat\Sigma_{il}\hat\Sigma_{jk}) & \hspace{-2mm}\text{if } \hat c_{ijkl}>0\\
    0 & \hspace{-2mm}\text{otherwise}
    \end{array}
    \right.\hspace{-8mm}
\end{equation}}
for $(i,j)$, $(k,l)\in\bar U$,  $\hat\Sigma_{ij}$ is the observed sample covariance (Equation~\eqref{eq:obscov}), and $\hat c_{ijkl}$, $\hat J$, and $\hat F$ are as in part (ii). 
\end{enumerate}
\end{corollary}
According to Corollary~\ref{coro:psihat}, the approximate covariance matrix $\Psi/n$ of the measurement error $\Delta$ (Equation~\eqref{eq:asympnormDelta}) can be estimated by $\hat\Psi/n$ (Equation~\eqref{eq:psiemphat}) and, if the data are Gaussian, by $\tilde\Psi/n$  (Equation~\eqref{eq:psigausshat}). Therefore, these estimates can be used in place of $\Phi$ to implement \glsname{} (Appendix~\ref{app:gls}). 
We compare the accuracy of $\hat\Psi$ and $\tilde\Psi$ via simulation in Appendix~\ref{app:veriftheo}. \glsname{} is implemented with $\tilde\Psi$ in the simulations of Section~\ref{sec:simulations}.

\subsection{Tuning parameter selection}\label{sec:tuningparameter}
We select the tuning parameter $\alpha$ in Equation~\eqref{eq:COValpha} via \textit{N-fold cross-validation} ($N$-CV), which we implement in Algorithm~\ref{algo:cv}. Let $\mathbf{X}^{(1)},...,\mathbf{X}^{(K)}$ be $K$ data sets about the observed node sets $V_1,...,V_K\subseteq V$. In step 1, we randomly split each data set $\mathbf{X}^{(j)}$ into $N$ subsets $\mathbf{X}^{(j)}_1,\ldots, \mathbf{X}^{(j)}_N$ of approximately equal sample size. We define the $h$-th fold as $\mathcal{X}_h=\{\mathbf{X}^{(1)}_h,\ldots, \mathbf{X}^{(K)}_h\}$. In step 2, for each fold $h=1,...,N$, we compute $\hat\Sigma_O^{(h)}$ from $\mathcal{X}_h$ and $\hat\Sigma_O^{(-h)}$ from $\{\mathcal{X}_l\}_{l\neq h}$, obtain a complete \methname{} covariance matrix $\hat\Sigma^{(-h)}(\alpha)$ based on $\hat\Sigma_O^{(-h)}$ via Algorithm~\ref{algo:algo1}, and compute the loss between $\hat\Sigma^{(-h)}_O(\alpha)$ and $\hat\Sigma^{(h)}_O$. In our analyses we use the squared loss between correlations in $O$: ${\rm loss}(\hat\Sigma^{(-h)}_O(\alpha),\hat\Sigma^{(h)}_O)=\sum_{(i,j)\in O}(\hat C^{(-h)}_{ij}(\alpha)-\hat C^{(h)}_{ij})^2 $. We only compute distances over the set $O$ primarily because the final estimates in $O^c$ are unaffected by $\alpha$. We finally average the $N$ losses to obtain $\textsc{Risk}_{\rm cv}(\alpha)$. The optimal $\alpha_{\rm cv}$ selected via $N$-CV is the minimizer
\begin{equation}\label{eq:alphaCV}
   \alpha_{\rm cv} ~:=~ \underset{0\le\alpha\le 1}{\arg\min} ~\textsc{Risk}_{\rm cv}(\alpha) 
\end{equation}
Besides selecting $\alpha$, $N$-CV can also be used to select among different baseline regression models $\mathcal{F}=\{f_1,\ldots,f_m\}$ in Algorithm~\ref{algo:algo1}. In this case, the model selection problem is augmented as
\begin{equation}
(\alpha_{\rm cv},f_{\rm cv}) ~=~ 
\underset{0\le\alpha\le 1,f\in\mathcal{F}}{\arg\min} ~\textsc{Risk}_{\rm cv}(\alpha,f),
\end{equation}
where $(\alpha_{\rm cv},f_{\rm cv})$ jointly minimizes the risk. We evaluate the performance of $N$-CV via simulation in Section~\ref{sec:simCV}.

\begin{algorithm}[ht] \small
\scbf{Input}: Incomplete data consisting of $K$ data sets $\mathbf{X}^{(1)},...,\mathbf{X}^{(K)}$ about the observed node sets $V_1,...,V_K\subseteq V$; $q$-dimensional auxiliary vectors $\{W_{ij}\}_{1\le i<j\le p}$; tuning parameter $\alpha \in [0,1]$; regression model $f:\mathbb{R}^q\to\mathbb{R}$; invertible function $g:(-1,1)\to\mathbb{R}$; loss function ${\rm loss}(x,y)$.\;
~

 \begin{enumerate}
    \item Randomly split each data set $\mathbf{X}^{(j)}$ into $N$ subsets $\mathbf{X}^{(j)}_1,\ldots, \mathbf{X}^{(j)}_N$ of approximately equal sample size. Define folds $\mathcal{X}_1,\ldots, \mathcal{X}_N$ where $\mathcal{X}_h=\{\mathbf{X}^{(1)}_h,\ldots, \mathbf{X}^{(K)}_h\}$. 
    \item For $h =1,...,N$
    \begin{enumerate}
        \item Estimate $\hat\Sigma_O^{(h)}$ from $\mathcal{X}_h$ and $\hat\Sigma_O^{(-h)}$ from $\{\mathcal{X}_l\}_{l\neq h}$.
        \item Obtain a complete covariance matrix estimate $\hat\Sigma^{(-h)}(\alpha)$ by using Algorithm~\ref{algo:algo1} with input $\hat\Sigma_O^{(-h)}$.
        \item Compute 
        $\textsc{loss}_h={\rm loss}(\hat\Sigma_O^{(-h)}(\alpha),\hat\Sigma_O^{(h)})$.
    \end{enumerate}
\end{enumerate}
\scbf{Output}: $N$-fold cross-validation risk 
\begin{equation}
    \textsc{Risk}_{\rm cv}(\alpha) ~=~ \frac{1}{N}\sum_{h=1}^N \textsc{loss}_h
\end{equation}
\caption{\methname{} $N$-fold Cross-Validation}\label{algo:cv}
\end{algorithm}

\subsection{Uncertainty quantification}\label{sec:auxcovUNCERT}
We use the bootstrap \citep{efron1994introduction,efron1994missing} to approximate the standard errors of the entries of the \methname{} estimator $\hat\Sigma(\alpha)$. For a given  function $\phi:\mathbb{R}^{p\times p}\to\mathbb{R}$, e.g., $\phi(\Sigma)=\Sigma_{12}$, the nonparametric bootstrap (Algorithm~\ref{algo:nonparbootstrap}) approximates the standard error of the estimator $\hat\theta=\phi(\hat\Sigma(\alpha_{\rm cv}))$ with the empirical standard deviation of multiple estimates $\tilde\theta_1,\ldots,\tilde\theta_B$ obtained from $B$ artificial data sets drawn with replacement from the original data, specifically by sampling with replacement from each data set $\mathbf{X}^{(1)},\ldots,\mathbf{X}^{(K)}$. The parametric bootstrap is similar to the nonparametric bootstrap except that the artificial data sets are generated from the estimated parametric distribution of the data. Assuming the data are Gaussian, in Algorithm~\ref{algo:parbootstrap} we simply generate $K$ independent datasets $\tilde{\mathbf{X}}^{(1)},\ldots,\tilde{\mathbf{X}}^{(K)}$, where $\tilde{\mathbf{X}}^{(k)}$ consists of $n_k$ i.i.d. samples drawn from $N(0,\hat\Sigma(\alpha_{\rm cv})_{V_kV_k})$. In Section~\ref{sec:simCV}, we show via simulation that both nonparametric and parametric bootstrap allow us to adequately approximate the true standard errors of the \methname{} covariance estimators.

\begin{algorithm}[ht]\small
{\bf Input}: Data sets $\mathbf{X}^{(1)},\ldots,\mathbf{X}^{(K)}$ with sample sizes $n_1,\ldots,n_K$, respectively; $V_1,\ldots,V_K$; function $\phi:\mathbb{R}^{p\times p}\to\mathbb{R}$; number of repeats $B$.\;

 For $b=1,\ldots,B$:
\begin{enumerate}
    \item Generate $K$ independent datasets $\tilde{\mathbf{X}}^{(1)},\ldots,\tilde{\mathbf{X}}^{(K)}$, where $\tilde{\mathbf{X}}^{(k)}$ consists of $n_k$ samples drawn\\ with replacement from $\mathbf{X}^{(k)}$.
    \item Based on $\tilde{\mathbf{X}}^{(1)},\ldots,\tilde{\mathbf{X}}^{(K)}$, find $\alpha_{\rm cv}$ via Algorithm~\ref{algo:cv} and compute $\tilde\Sigma(\alpha_{\rm cv})$ via Algorithm~\ref{algo:algo1}.
    \item Compute $\tilde\theta_b=\phi(\tilde\Sigma(\alpha_{\rm cv}))$.
\end{enumerate}
{\bf Output}: Estimate of the standard error of $\hat\theta=\phi(\hat\Sigma(\alpha_{\rm cv}))$:
\begin{equation}
\widehat{se}(\hat\theta) = \left(\tfrac{1}{B-1}\sum_{b=1}^B \left(\tilde\theta_b-\tfrac{1}{B}\sum_{b=1}^B\tilde\theta_b\right)^2\right)^{1/2}
\end{equation}
\caption{\methname{} Nonparametric bootstrap}\label{algo:nonparbootstrap}
\end{algorithm}

\begin{algorithm}[ht]\small
{\bf Input}: \methname{} estimate $\hat\Sigma(\alpha_{\rm cv})$; $V_1,\ldots,V_K$; sample sizes $n_1,\ldots,n_K$ of the $K$ observed data $\mathbf{X}^{(1)},\ldots,\mathbf{X}^{(K)}$; function $\phi:\mathbb{R}^{p\times p}\to\mathbb{R}$; number of repeats $B$.\;

For $b=1,\ldots,B$:
\begin{enumerate}
    \item Generate $K$ independent datasets $\tilde{\mathbf{X}}^{(1)},\ldots,\tilde{\mathbf{X}}^{(K)}$, where $\tilde{\mathbf{X}}^{(k)}$ consists of $n_k$ i.i.d. samples\\ drawn from $N(0,\hat\Sigma(\alpha_{\rm cv})_{V_kV_k})$.
    \item Based on $\tilde{\mathbf{X}}^{(1)},\ldots,\tilde{\mathbf{X}}^{(K)}$, find $\alpha_{\rm cv}$ via Algorithm~\ref{algo:cv} and compute $\tilde\Sigma(\alpha_{\rm cv})$ via Algorithm~\ref{algo:algo1}.
    \item Compute $\tilde\theta_b=\phi(\tilde\Sigma(\alpha_{\rm cv}))$.
\end{enumerate}
{\bf Output}: Estimate of the standard error of $\hat\theta=\phi(\hat\Sigma(\alpha_{\rm cv}))$:
\begin{equation}
\widehat{se}(\hat\theta) = \left(\tfrac{1}{B-1}\sum_{b=1}^B \left(\tilde\theta_b-\tfrac{1}{B}\sum_{b=1}^B\tilde\theta_b\right)^2\right)^{1/2}
\end{equation}
\caption{\methname{} (Gaussian) Parametric bootstrap}\label{algo:parbootstrap}
\end{algorithm}

\section{Simulations}\label{sec:simulations}
In this section, we present the results of an extensive simulation study to assess the performance of \methname{} in various settings. In Section~\ref{sec:simGround}, we describe how our ground truth covariance matrices are constructed and how the incomplete data are generated. In Section~\ref{sec:simCV}, we assess the accuracy of our proposed \methname{} cross-validation procedure (Section~\ref{sec:tuningparameter}) at selecting the optimal parameter  $\alpha$, and also at selecting the baseline regression model $f$. In Section~\ref{sec:simuncert}, we investigate the accuracy of the bootstrap (Section~\ref{sec:auxcovUNCERT}) at approximating the standard errors of the \methname{} covariance estimates. Finally, in Section~\ref{sec:simCompare}, we compare the performance of \methname{} with other   covariance matrix completion methods.

\subsection{Simulation settings}\label{sec:simGround}

We generate our ground truth covariance matrices as follows. First, we generate a correlation matrix $C$, where $C_{ii}=1$ for all $i=1,\ldots,p$, and
\begin{equation}\label{eq:gamma}
    C_{ij} = \sqrt{\tfrac{\gamma}{2}} W_{ij}+\sqrt{\tfrac{1-\gamma}{2}}Z_{ij},~~ 1 \leq i < j \leq p
\end{equation}
where $W_{ij},Z_{ij}, 1 \leq i < j \leq p$, are i.i.d. ${\rm Uniform}(-1,1)$, and the parameter $\gamma\in [0,1]$ regulates the relative importance of the observed auxiliary variable $W_{ij}$ compared with the noise component $Z_{ij}$. It is easy to verify that $-1<C_{ij}<1$ almost surely, and $\var(C_{ij})=\frac{1}{6}$ for all $\gamma\in[0,1]$, i.e., $\var(C_{ij})$ is independent of $\gamma$, which helps reduce undesired side effects on our simulation results due to using different values of $\gamma$. Finally, we obtain $\Sigma$ as the positive definite corrected version of $C$ via Algorithm~\ref{algo:posdiagadd}, Appendix~\ref{app:poscorrection}.

We generate Gaussian random vectors $X^{(1)},...,X^{(n)}\iid N(0,\Sigma)$ and then drop data to produce the desired proportion of missingness $\eta$ (Equation~\eqref{eq:eta}). For example, if $K=2$ and, for simplicity, $n$ is even, we split the data into two subsets $X^{(1)},\ldots,X^{(n/2)}$ and $X^{(n/2+1)},\ldots,X^{(n)}$, and then obliterate all values about $V_1^C$ from the first data subset and all values about $V_2^C$ from the second data subset. We set $V_1=1,...,p-s$ and $V_2=s+1,...,p$ and take $s$ to achieve the desired proportion of missingness $\eta$.

\subsection{Accuracy of tuning parameter selection}\label{sec:simCV}
In this simulation, we investigate the performance of 10-fold cross-validation (Algorithm~\ref{algo:cv}) at selecting the tuning parameter $\alpha$ and the regression model $f$ used in Algorithm~\ref{algo:algo1}. In Section~\ref{sec:simCV1}, we postulate that $f$ possesses a known linear form (\olsname, Section~\ref{sec:naive}), while in  Section~\ref{sec:simCV2} we do not impose this assumption and estimate $f$ nonparametrically via splines (\splinesname, Section~\ref{sec:naive}). In Appendix~\ref{app:simfgls}, we further investigate tuning parameter selection when $f$ is fitted linearly while accounting for measurement error (\glsname, Section~\ref{sec:fgls}).

\subsubsection{Known regression model}\label{sec:simCV1}
\begin{figure}[t!]
    {\footnotesize\sf (a)}\vspace{-3mm}

\includegraphics[width=1\columnwidth]{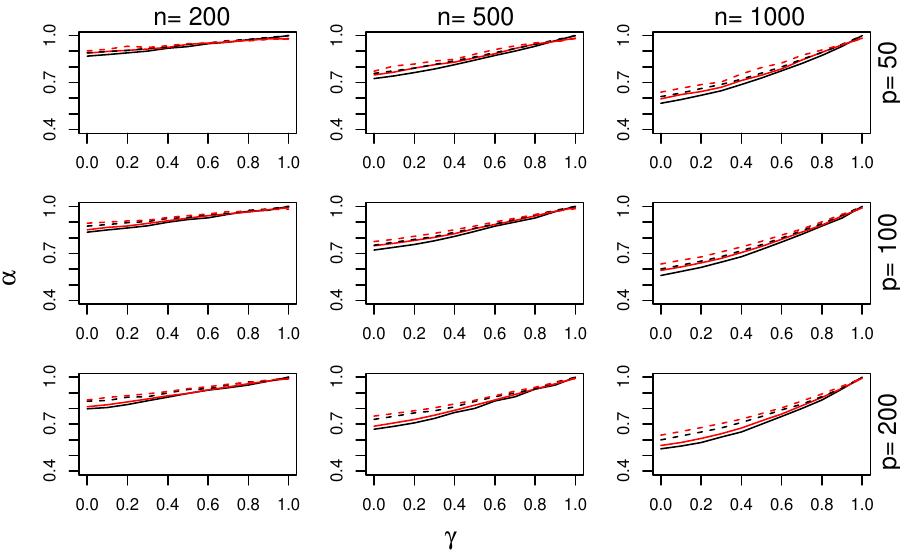}\vspace{-5mm}\\

{\footnotesize\sf (b)}\vspace{-3mm}

\includegraphics[width=1\columnwidth]{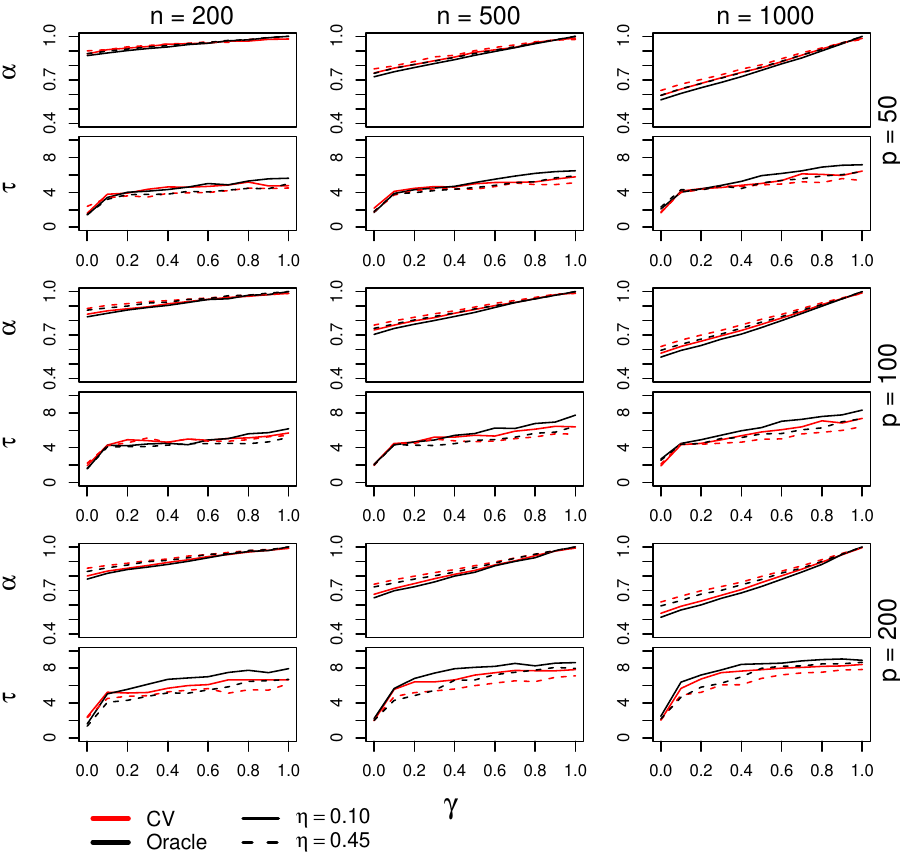}

\caption{{\bf Tuning parameters selection via 10-CV}. {\bf (a)} Average tuning parameter $\alpha_{\rm cv}$ selected in \olsname{} via $10$-CV (red) and oracle counterpart $\alpha_{\rm or}$ (black) are plotted against $\gamma$ (Equation~\eqref{eq:gamma}), for various scenarios with number of data sets $K=2$, missingness proportion $\eta=0.1,0.45$, number of variables $p=50,100,200$, and total sample size $n=200,500,1000$. The average $\alpha_{\rm cv}$ and $\alpha_{\rm or}$ are very close, increase with $\gamma$ and $\eta$, decrease with $n$, and are weakly affected by $p$. 
{\bf (b)} Average tuning parameters $\alpha_{\rm cv}$ and $\tau_{\rm cv}$ selected in \splinesname{} via $10$-CV (red) and oracle counterparts $\alpha_{\rm or}$ and $\tau_{\rm or}$ (black) are plotted against $\gamma$  (Equation~\eqref{eq:gamma}), for the same scenarios as in (a). The average $\alpha_{\rm cv}$ and $\alpha_{\rm or}$ are very close and both increase with $\gamma$ and $\eta$, decrease with $n$, and are weakly affected by $p$. The average $\tau_{\rm cv}$ and $\tau_{\rm or}$ are very close and both increase with $\gamma$, $n$, and $p$, and decrease with $\eta$.}
    \label{fig:CV v Oracle}
\end{figure}

We generate ground truth covariance matrix and data as described in Section~\ref{sec:simGround} in various settings: $\gamma\in [0,1]$, $p=50,100,200$, $n=200,500,1000$, $K=2$, and $\eta=0.1,0.45$. Equation~\eqref{eq:gamma} implies a linear relationship between correlation $C_{ij}$ and auxiliary variable $W_{ij}$, and in this simulation we assume this to be known. We implement \olsname{} (Section~\ref{sec:naive}) assuming $f(w)=\beta_0+\beta_1w$, and select $\alpha\in[0,1]$ via $10$-CV. We compare the selected $\alpha_{\rm cv}$ with the {\it oracle tuning parameter}
\begin{equation}
    \alpha_{\rm or} ~:=~ \underset{0\le \alpha\le 1}{\arg\min}~ \sum_{(i,j)\in O}(\hat C_{ij}(\alpha)-C_{ij})^2,
\end{equation}
where $C_{ij}$ is the ground truth correlation between variables $(i,j)$, and $\hat C_{ij}(\alpha)$ is the correlation estimate yielded by Algorithm~\ref{algo:algo1} with tuning parameter $\alpha$. For all simulation settings, we compute the average $\alpha_{\rm cv}$ and the average $\alpha_{\rm or}$ across $B=100$ simulation repeats. We summarize the results in Figure~\ref{fig:CV v Oracle}(a). We can see that $\alpha_{\rm cv}$ adequately recovers $\alpha_{\rm or}$, suggesting that our cross-validation procedure is likely to yield a covariance matrix estimate $\hat\Sigma(\alpha_{\rm cv})$ that is close to the oracle optimal $\hat\Sigma(\alpha_{\rm or})$. 
Moreover, $\alpha_{\rm cv}$ and $\alpha_{\rm or}$ both increase with $\gamma$, likely because a larger value of this parameter strengthens the relationship between the correlations and the auxiliary variable, so the auxiliary baseline covariances are assigned a larger weight than the empirical covariances. Note that $\alpha_{\rm cv}$ and $\alpha_{\rm or}$ do not necessarily approach $0$ as $\gamma$ approaches $0$ presumably because, even when the auxiliary variables are uninformative, shrinking the correlation estimates towards an average correlation reduces estimation variability. 
Furthermore, $\alpha_{\rm cv}$ and $\alpha_{\rm or}$ decrease with $n$ presumably because a larger sample size makes empirical covariances more reliable in the observed set $O$. Similar results were obtained with \glsname{} (see Appendix~\ref{app:simfgls}).  On the other hand, a larger $\eta$ implies a smaller overlap between the two observed sets $V_1$ and $V_2$, and thereby smaller average sample sizes across the entries in $O$, yielding larger $\alpha_{\rm cv}$ and $\alpha_{\rm or}$. Finally, the number of variables $p$ appears to affect $\alpha_{\rm cv}$ and $\alpha_{\rm or}$ only slightly.  

\subsubsection{Unknown regression model}\label{sec:simCV2} 
We now assume that the regression model has an unknown form, potentially highly nonlinear. In this case, we fit \splinesname{} (Section~\ref{sec:naive}), where $f$ is fitted by splines with $\tau$ knots, and we evaluate the performance of $10$-CV at selecting simultaneously the optimal $\alpha\in[0,1]$ and $\tau\in\mathcal{T}$, where $\mathcal{T}$ is some reasonable set of integers. We generate the ground truth covariance matrix and the data with the same settings as in Section~\ref{sec:simCV1} except that, in Equation~\eqref{eq:gamma}, we replace $W_{ij}$ with $\sin(7 W_{ij})$, implying a \textit{nonlinear} relationship between the correlation $C_{ij}$ and the auxiliary variable $W_{ij}$. We compare the jointly selected $(\alpha_{\rm cv},\tau_{\rm cv})$ with the \textit{joint oracle parameters}
\begin{equation}
    (\alpha_{\rm or},\tau_{\rm or}) := \underset{0\le \alpha\le 1,~\tau\in\mathcal{T}}{\arg\min} \sum_{(i,j)\in O}(\hat C_{ij}(\alpha,\tau)-C_{ij})^2,
\end{equation}
where $\hat C_{ij}(\alpha,\tau)$ is the correlation estimate yielded by Algorithm~\ref{algo:algo1} with tuning parameter $\alpha$ and $f$ estimated using splines with $\tau$ knots. 
In Figure~\ref{fig:CV v Oracle}(b), we can see that the averages $(\alpha_{\rm cv},\tau_{\rm cv})$ are very close to $(\alpha_{\rm or},\tau_{\rm or})$. The selected values $\alpha_{\rm cv}$ and $\alpha_{\rm or}$ behave similarly to the case where the regression model is prespecified (Figure~\ref{fig:CV v Oracle}(a); Section~\ref{sec:simCV1}). The numbers of knots $\tau_{\rm cv}$ and $ \tau_{\rm or}$ both increase with $\gamma$, $p$, and $n$, presumably because the nonlinear features of the relationship between correlation and the auxiliary variable become easier to detect when $\gamma$, $p$, and $n$ are larger. On the other hand, a larger $\eta$ reduces the sample size used in the spline regression fit, favouring the selection of simpler models with a smaller number of knots.
\begin{figure}[ht]
\includegraphics[width=1\columnwidth]{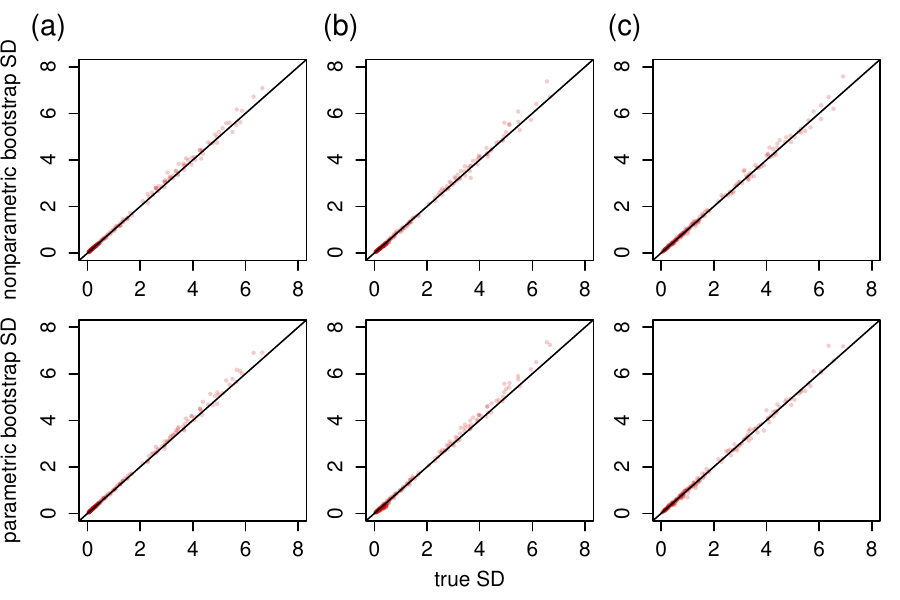}\vspace{-4mm}
    \caption{{\bf Bootstrap}. Nonparametric (top) and parametric (bottom) bootstrap standard errors of the entries of the covariance matrix estimators \textbf{(a)} \olsname{}, \textbf{(b)} \glsname{}, and \textbf{(c)} \splinesname{}  plotted against the true standard errors approximated via Monte Carlo integration, for the case $p=20$, $n=5000$, $K=2$, $\eta=0.3$, and $\gamma=0.8$. 
    }
    \label{fig:bootstrapsim}
\end{figure}

\subsection{Accuracy of uncertainty quantification}\label{sec:simuncert}

In this simulation, we assess the accuracy of the nonparametric bootstrap (Algorithm~\ref{algo:nonparbootstrap}) and the parametric bootstrap (Algorithm~\ref{algo:parbootstrap}) at approximating the standard errors of the entries of the covariance estimators \olsname{} (Section~\ref{sec:naive}), \glsname{} (Section~\ref{sec:fgls}), and \splinesname{} (Section~\ref{sec:naive}). In Figure~\ref{fig:bootstrapsim}, we summarize the results of a simulation with $p=20$, $n=5000$, $K=2$, $\eta=0.3$, $\gamma=0.8$, and number of bootstrap samples $B=500$; the data were generated as in  Section~\ref{sec:simCV1} for \olsname{} and \glsname{}, and as in Section~\ref{sec:simCV2} for \splinesname{}. We can see that both nonparametric and parametric bootstrap procedures let us adequately estimate the true standard errors, which we approximated via Monte Carlo integration (2000 draws).

\subsection{Methods comparison}\label{sec:simCompare}
In this simulation, we compare the performance of \methname{} with two alternative covariance matrix completion methods:
\begin{enumerate}[(i).]
\item \textsc{\textbf{MaxDet}}.~ The \textit{max-determinant} positive definite completion \citep{grone1984positive} of $\hat\Sigma_O$ is
    \begin{equation}\label{eq:MaxDet}
        \hat\Sigma_{\rm MaxDet} = \underset{S\succ 0,S_O=\hat\Sigma_O}{\arg\max} ~\det S
    \end{equation}
Closed-form solutions of this optimisation problem are given in \cite{georgescu2018explicit}. 
\item \textbf{LR}.~ In this \textit{low-rank} approach, we first apply low-rank matrix completion  \citep{candes2009exact} to obtain a completed version of the partially observed data matrix $\tilde{\mathbf{X}}$ given by the union of $K$ data sets $\mathbf{X}_1,\ldots,\mathbf{X}_K$, 
\begin{eqnarray}\label{eq:LR}
&&\mathbf{X}^* = \underset{\mathbf{Z}\in\mathbb{R}^{n\times p}}{\arg\min} ~{\rm rank}(\mathbf{Z})\\
&& {\rm s.t.} ~ \mathbf{Z}_\Omega=\tilde{\mathbf{X}}_\Omega\nonumber 
\end{eqnarray}
where $\Omega$ is the set of observed data entries. Then, we obtain a complete covariance matrix estimate $\hat\Sigma_{\rm LR}$ as the sample covariance matrix of $\mathbf{X}^*$. We solve the optimisation problem in Equation~\eqref{eq:LR} by using the algorithm proposed by \cite{mazumder2010spectral} and summarized in \cite{vinci2024unsupervised}. 
\end{enumerate}
We generate the ground truth covariance matrix and the data as described in Section~\ref{sec:simGround} with various settings: $\gamma\in[0,1]$, $p=50$, $n=500,1000$, $K=2$, $\eta=0.3$. For a given covariance matrix estimator $\hat\Sigma$, we evaluate the performance of each method in terms of average squared losses for correlations and partial correlations, as summarized in Table~\ref{tab:losses}. 

\begin{table}[ht]
\caption{{\bf Performance metrics}. Given an estimate $\hat\Sigma$ of $\Sigma$, we compute losses between correlations $\hat C_{ij}=\hat\Sigma_{ij}(\hat\Sigma_{ii}\hat\Sigma_{jj})^{-1/2} $ and $C_{ij}=\Sigma_{ij}(\Sigma_{ii}\Sigma_{jj})^{-1/2} $, and between partial correlations $\hat\rho_{ij}=-\hat\Theta_{ij}(\hat\Theta_{ii}\hat\Theta_{jj})^{-1/2} $ and $\rho_{ij}=-\Theta_{ij}(\Theta_{ii}\Theta_{jj})^{-1/2} $,  where $\Theta=\Sigma^{-1}$ and $\hat\Theta=\hat\Sigma^{-1}$.}\label{tab:losses}
{ \begin{tabular}{lcc}
\toprule
 &  $\mathbf{O}$ & $\mathbf{O^c}$ \\
\midrule
Correlation & \multicolumn{1}{|c|}{$\frac{\sum\limits_{i\neq j,(i,j)\in O} (\hat C_{ij}-C_{ij})^2}{|O|-p}$}    & ~~$\frac{\sum\limits_{(i,j)\in O^c} (\hat C_{ij}-C_{ij})^2}{|O^c|}$~~   \\
Partial Correl.  & \multicolumn{1}{|c|}{$\frac{\sum\limits_{i\neq j,(i,j)\in O} (\hat \rho_{ij}-\rho_{ij})^2}{|O|-p}$}  & ~~$\frac{\sum\limits_{(i,j)\in O^c} (\hat \rho_{ij}-\rho_{ij})^2}{|O^c|}$~~\\
\bottomrule
\end{tabular}}
\end{table}

\begin{figure}[t!]
    \includegraphics[width=1\columnwidth]{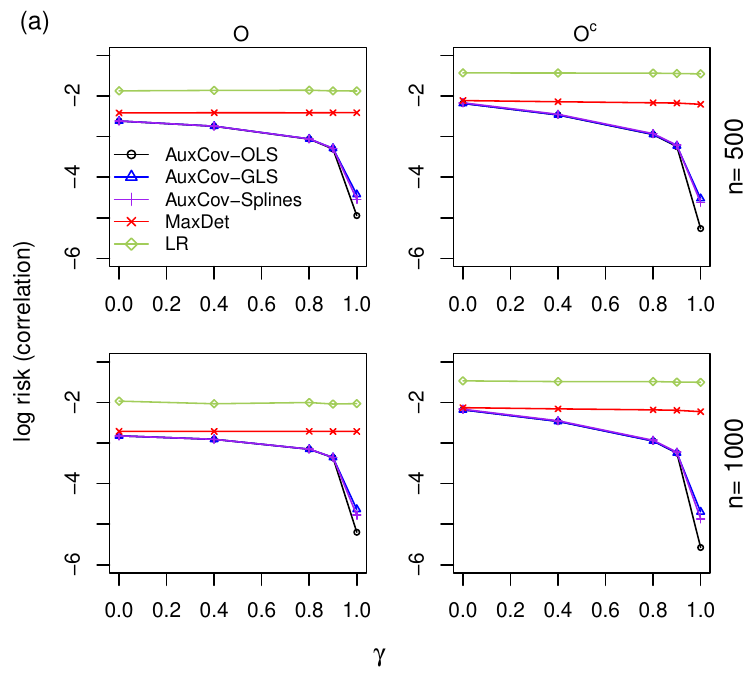}\\
    \includegraphics[width=1\columnwidth]{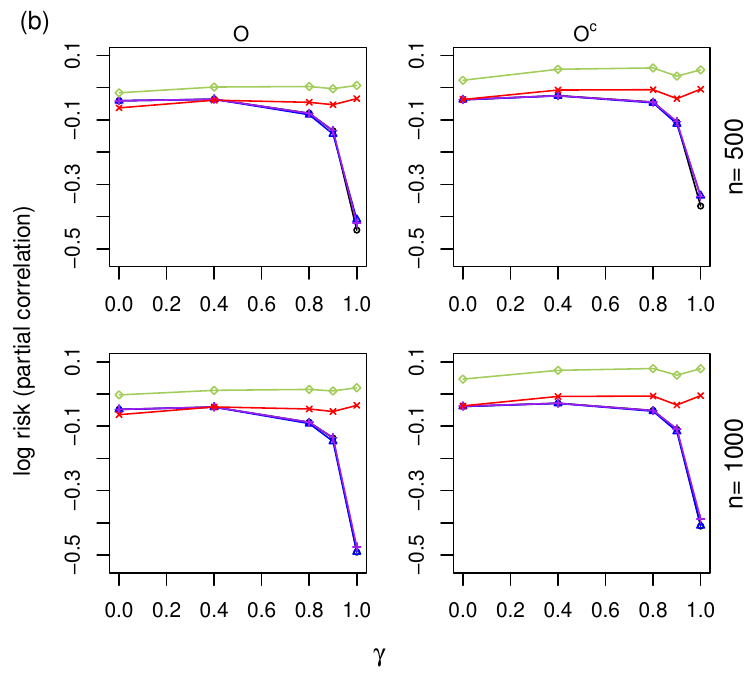}\vspace{-3mm}
   \caption{{\bf Methods comparison.} 
    Performance of \olsname{}, \glsname{}, \splinesname{}, MaxDet, and LR at recovering {\bf (a)} correlations and {\bf (b)} partial correlations for $\gamma\in[0,1]$, $p=50$, $n=500,1000$, $K=2$, $\eta=0.3$. All \methname{} variants outperform MaxDet and LR in almost all conditions, especially for larger values of $\gamma$.
    }
    \label{fig:comparemethodssim}
\end{figure}

In Figure~\ref{fig:comparemethodssim}, we show the log$_{10}$ losses in Table~\ref{tab:losses} averaged over 100 simulation repeats. We can see that all \methname{} variants outperform MaxDet and LR under almost all conditions, especially for larger values of $\gamma$. Additional simulation results are reported in Appendix~\ref{app:addsimmodelcomp}. 

\section{Estimating correlations from incomplete neuronal data}\label{sec:data}
Functional neuronal connectivity is the statistical dependence among neurons \citep{vinci2016separating,vinci2018adjusted,vinci2018adjustedcovariance}, and estimating the covariance matrix of neurons' activities is a fundamental step in the analysis of functional connectivity. The electrophysiological activity of thousands of neurons is now commonly recorded using cutting-edge technologies such as 2-photon calcium imaging \citep{stringer2019spontaneous,microns2021functional}. However, due to technological limitations, recording from a large population of neurons simultaneously with fine temporal resolution is often impractical. Thus, it is preferable to record from smaller subsets of neurons simultaneously, maintaining fine temporal resolution, but yielding data sets characterized by the structural missingness considered in this paper. 

We apply \methname{} to the analysis of neuronal calcium imaging activity traces data publicly available, \citep{stringer2019spontaneous} recorded from around ten thousand neurons in a 1mm $\times$ 1mm $\times$ 0.5mm portion of mouse visual cortex (70–385µm depth). The data were recorded \textit{in vivo} via 2-photon calcium imaging of GCaMP6s with 2.5Hz scan rate \citep{pachitariu2017suite2p}. During the experiment, the animal was free to run on an air-floating ball in complete darkness for 105 minutes. For our analyses, we focus on the most superficial layer (70 µm depth) of the brain portion consisting of 725 neurons. Calcium activity traces are summed in groups of three consecutive time bins and then square-rooted before computing sample covariances. 

To validate the performance of \methname{}, we first compute the \methname{} covariance matrix estimate using the full dataset ($\eta=0, n=5000$) and the auxiliary variable inter-neuron distance (mm). This estimate serves as the ground truth covariance matrix, which we then compare with \methname{} covariance estimates derived from incomplete data obtained by systematically removing parts of the data. We fit \methname{} first with a baseline regression estimated parametrically, and then with one fitted nonparametrically. In particular, we apply \olsname{}, which is the most computationally tractable given the large dimensionality $p=725$, and then implement \splinesname{}.

\begin{figure}[t!]

\includegraphics[width=1\columnwidth]{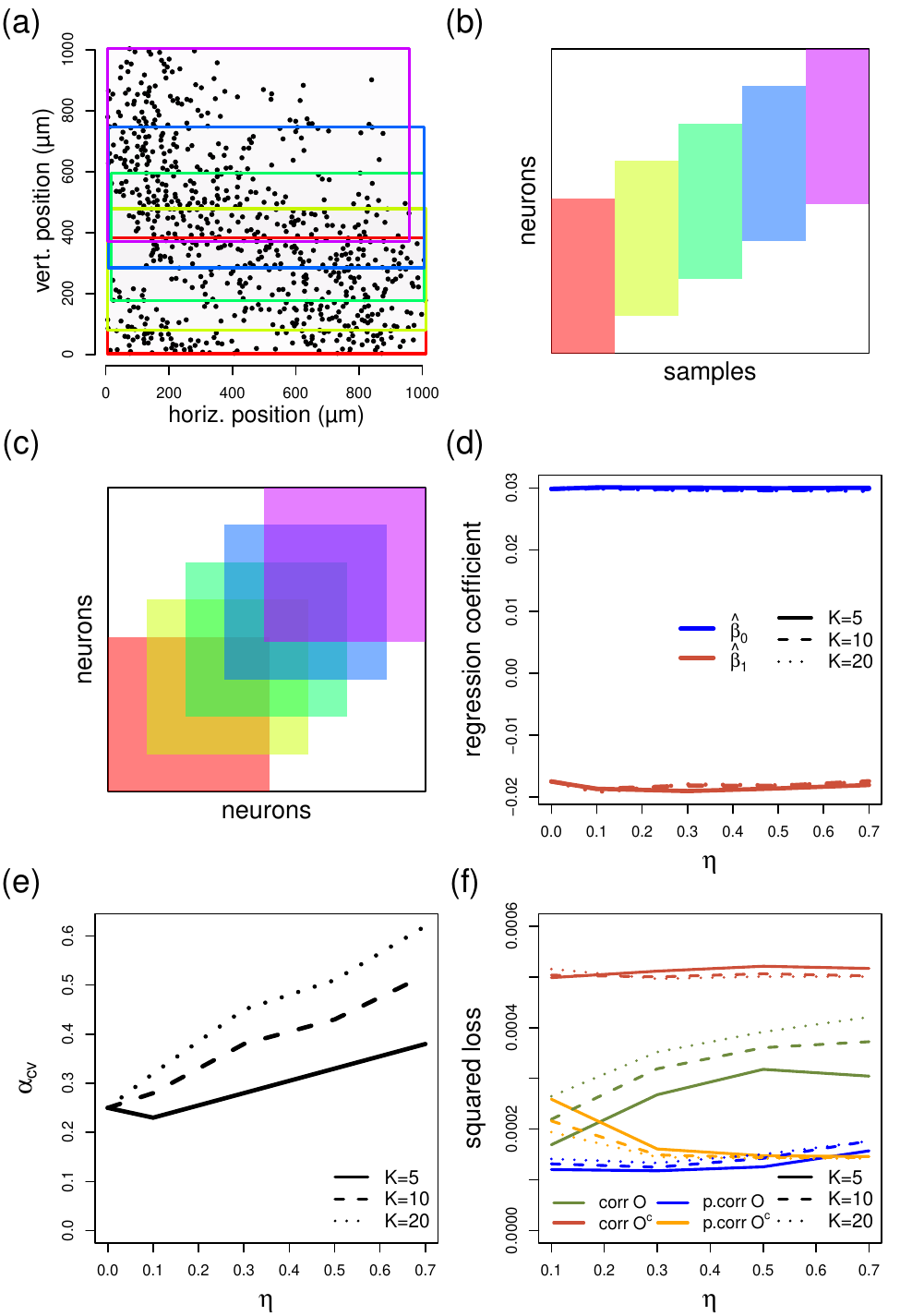}

\hfill\includegraphics[width=.985\columnwidth]{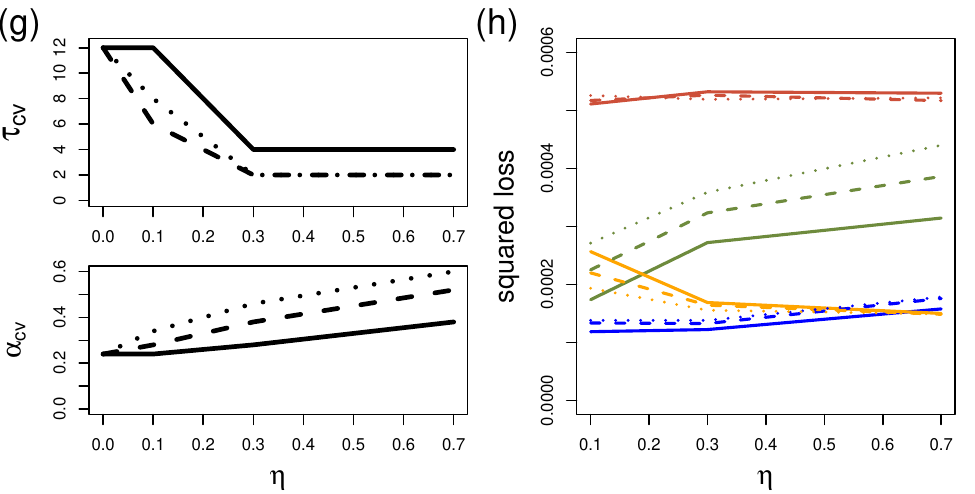}

\caption{{\bf Analysis of neuronal data.}
    {\bf (a)} Locations of 725 neurons and $K=5$ neurons subsets. 
    {\bf (b)} Incomplete data from observed neurons in (a). 
    {\bf (c)} Example of set of observed neuronal pairs ($\eta=0.3$). 
    {\bf (d)} Estimated coefficients of the \olsname{} baseline linear regression for various observation patterns characterized by number of variable subsets $K=5,10,20$ and pairwise missingness proportion $\eta\in[0,0.7]$. 
    {\bf (e)} \olsname{} tuning parameter $\alpha_{\rm cv}$ selected via 10-CV for $K=5,10,20$ and $\eta\in[0,0.7]$. 
    {\bf (f)} Squared loss of the \olsname{} correlation and partial correlation estimates in $O$ and in $O^c$ (Table~\ref{tab:losses}) for $K=5,10,20$ and $\eta\in[0.1,0.7]$.
    {\bf (g)} Number of splines knots $\tau_{\rm cv}$ and tuning parameter $\alpha_{\rm cv}$ selected in \splinesname{} via 10-CV for $K=5,10,20$ and $\eta\in[0,0.7]$. 
    {\bf (h)} Squared loss of the \splinesname{} correlation and partial correlation estimates in $O$ and in $O^c$ for $K=5,10,20$ and $\eta\in[0.1,0.7]$.}
    \label{fig:data}
\end{figure}

Figure \ref{fig:data}(a) depicts the positions of the 725 neurons. Additionally, it illustrates an observation scheme involving $K = 5$ subsets of recorded neurons, each delineated by a colored rectangle. Each subset of neurons is observed with a sample size $n_k=n/K$, as shown in Figure~\ref{fig:data}(b). The sizes of the neuronal subsets induce a missingness proportion $\eta$ (Equation~\eqref{eq:eta}) in the observed variable pairs (Equation~\eqref{eq:obscov}) depicted in Figure~\ref{fig:data}(c). In Figure~\ref{fig:data}(d), we show that the values of the estimated coefficients $\hat\beta_0$ (positive) and $\hat\beta_1$ (negative) of the linear baseline regression function $f(w)=\beta_0+\beta_1w $ of \olsname{} only slightly vary with $K=5,10,20$ and $\eta\in[0,0.7]$. In Figure~\ref{fig:data}(e), we can see that the \olsname{} tuning parameter $\alpha_{\rm cv}$ (Equation~\eqref{eq:alphaCV}) selected via $10$-CV (Algorithm~\ref{algo:cv}) increases with $\eta$, in agreement with the simulation results in Section~\ref{sec:simCV1}, and also with $K$, suggesting an increasing importance of the baseline correlation matrix (Equation~\eqref{eq:corrbaseline}) for more intricate scenarios of structural missingness. In Figure~\ref{fig:data}(f), we plot the values of the four loss functions defined in Table~\ref{tab:losses} 
versus $\eta\in [0.1,0.7]$ and for $K=5,10,20$, using the full data ($\eta=0$) \olsname{} covariance matrix estimate as the ground truth covariance matrix. We can see that the performance of \olsname{} is robust to different levels of missingness and observation patterns. 

Similar performance was observed using \splinesname{}. In Figure~\ref{fig:data}(g), we show the number of spline knots $\tau_{\rm cv}$ and tuning parameter $\alpha_{\rm cv}$ of \splinesname{} selected via 10-CV (Algorithm~\ref{algo:cv}). We observe that $\tau_{\rm cv}$ decreases with $\eta$, while $\alpha_{\rm cv}$ increases with $\eta$, in agreement with the simulation results in Section~\ref{sec:simCV2}. Finally, in Figure~\ref{fig:data}(h), we can see that the performance of \splinesname{} evaluated in terms of squared losses (Table~\ref{tab:losses}) is robust to different levels of missingness and observation patterns.

\section{Discussion and conclusion}\label{sec:disc}

In this article, we proposed a novel and flexible approach, \methname{}, for the completion and regularization of covariance matrix estimates by leveraging auxiliary variables. 

Auxiliary variables are supplementary quantities that are available beyond the observed traditional data and that may be informative about the unknown parameters of a statistical model. \methname{} first fits the relationship between the observed correlation estimates and the auxiliary variables through regression, and then uses it to predict the missing covariance matrix entries. Furthermore, the information extracted from the auxiliary variables is also used to improve the observed covariance estimates. 

We explored various approaches to fitting the \methname{} baseline regression function, including linear regression and nonparametric regression. We proposed a cross-validation procedure for the selection of the \methname{} tuning parameters, and bootstrap algorithms to quantify the uncertainty of the estimator. We evaluated the performance of \methname{} in simulations, where we observed that \methname{} can substantially outperform methods that do not exploit auxiliary information, and in the analysis of neuroscience data, where we used the physical distance between neurons as an auxiliary variable.

\methname{} can be applied to the completion of generic incomplete covariance matrix estimators, and we expect it to be useful in the analysis of large-scale multivariate data from disparate research fields. 
We plan to extend \methname{} in multiple directions, including graphical modelling, the Bayesian framework, and regularized regression in the case of large numbers of auxiliary variables.

\section*{Acknowledgements}

\noindent {\bf Funding}~ J.S. was supported by the Arthur J. Schmitt Presidential Leadership Fellowship.
\\
{\bf Data availability}~ The neuroscience data set \citep{stringer2019spontaneous} used in our analyses is publicly available at  {\small \url{https://figshare.com/articles/dataset/Recordings_of_ten_thousand_neurons_in_visual_cortex_during_spontaneous_behaviors/6163622}}, file name ``spont\_M161025\_MP030\_2016-11-20.mat.''

\appendix
\section{Algorithms}
\subsection{Positive Definite Matrix Correction}\label{app:poscorrection}
Let $A$ be a $p\times p$ symmetric matrix with positive diagonals but with nonpositive minimum eigenvalue $\lambda_{\rm min}(A)\le 0$. A positive definite corrected version of $A$ is given by
\begin{equation}
\tilde A= {\rm diag}(A)^{\frac{1}{2}} {\rm diag}(\tilde B)^{-\frac{1}{2}} \tilde B {\rm diag}(\tilde B)^{-\frac{1}{2}}{\rm diag}(A)^{\frac{1}{2}},
\end{equation}
where 
\begin{eqnarray}
\tilde B &=& B+\nu^* I_p,\\
B&=& {\rm diag}(A)^{-\frac{1}{2}} A ~{\rm diag}(A)^{-\frac{1}{2}},
\end{eqnarray}
and
\begin{eqnarray}
   \left\{
  \begin{array}{c} \nu^*=\underset{\nu>0}{\arg\min}~ \nu\\
\text{s.t.}~B+\nu I_p\succ 0 
\end{array}
\right.
\end{eqnarray}
This approach can be implemented via Algorithm~\ref{algo:posdiagadd}, where $\delta$ is a small positive constant, e.g., $\delta=0.001$.

\begin{algorithm}[h]\small
\scbf{Input}: Symmetric $p\times p$ matrix $A$ with positive diagonals; positive constant $\delta>0$.\;~\\
 \begin{enumerate}
    \item Compute $B = {\rm diag}(A)^{-\frac{1}{2}} A {\rm diag}(A)^{-\frac{1}{2}}$
    \item \textbf{While} $\lambda_{\rm min}(B)\le 0$:\\ 
    Update $B\equiv B+\delta I_p$
\end{enumerate}

~

\scbf{Output}: Positive definite matrix
\[\tilde A ~=~ {\rm diag}(A)^{\frac{1}{2}}{\rm diag}(B)^{-\frac{1}{2}} B{\rm diag}(B)^{-\frac{1}{2}} {\rm diag}(A)^{\frac{1}{2}}\]
\caption{Positive Definite Matrix Correction}\label{algo:posdiagadd}
\end{algorithm}

\begin{algorithm}[ht]\small
\scbf{Input}: Response vector $y\in\mathbb{R}^m$; design matrix $\mathcal{W}\in\mathbb{R}^{m\times (q+1)}$; $m\times m$ positive definite covariance matrix $\Phi$; start values $\hat\beta\in\mathbb{R}^{q+1}$ and $\hat\varphi=2\log\hat\sigma_\varepsilon\in\mathbb{R}$; step size $b>0$; acceleration parameter $a>0$; stopping threshold $s$; $t=0$; maximum number of iterations $T>0$.\;

    ~
    
    ~\textbf{While} $t<T$:
    \begin{enumerate}
        \item Compute the gradient \\ $\nabla\ell(\hat\beta,\hat\varphi)=( \frac{\partial\ell}{\partial\beta}(\hat\beta,\hat\varphi),\frac{\partial\ell}{\partial\beta}(\hat\beta,\hat\varphi))^T$, where
        \begin{eqnarray*}
            \tfrac{\partial\ell}{\partial\beta}(\hat\beta,\hat\varphi) &=& 2\mathcal{W}^{T}(e^{\hat\varphi}I + \Phi)^{-1}(y -\mathcal{W}\hat\beta)\\ 
            \tfrac{\partial\ell}{\partial\varphi}(\hat\beta,\hat\varphi) &=& -e^{\hat\varphi}\text{tr}((e^{\hat\varphi}I + \Phi)^{-1})\\
            && \hspace{-2mm}+ e^{\hat\varphi}(y - \mathcal{W}\hat\beta)^T
            (e^{\hat\varphi}I + \Phi)^{-2}
            (y - \mathcal{W}\hat\beta)
        \end{eqnarray*}
        \item Compute $(\tilde\beta,\tilde\varphi)^T = (\hat\beta,\hat\varphi)^T + b\cdot \nabla\ell(\hat\beta,\hat\varphi)$.
        \item If $\ell(\tilde\beta,\tilde\varphi) \le \ell(\hat\beta,\hat\varphi)$, update $b := b/a$ and return to step 2.
        \item If $\ell(\tilde\beta,\tilde\varphi) > \ell(\hat\beta,\hat\varphi)$ and $\left\vert \frac{\ell(\tilde\beta,\tilde\varphi) - \ell(\hat\beta,\hat\varphi)}{\ell(\hat\beta,\hat\varphi)}\right\vert < s$, update $(\hat\beta,\hat\varphi) := (\tilde\beta, \tilde\varphi)$ and \textbf{stop}.
        \item If $\ell(\tilde\beta,\tilde\varphi) > \ell(\hat\beta,\hat\varphi)$, 
        update $(\hat\beta,\hat\varphi) := (\tilde\beta, \tilde\varphi)$, $b := b\cdot a$, and $t:=t+1$. 
    \end{enumerate}
~

\scbf{Output}: Estimates $(\hat\beta,\hat\sigma_\varepsilon^2)=(\hat\beta,e^{\hat\varphi})$.
\caption{Gradient Ascent for GLS regression}\label{algo:optbaselinemeaserr}
\end{algorithm}

\subsection{Gradient ascent for \methname{}-GLS}\label{app:gls}
To obtain the MLE of $(\beta,\sigma_\varepsilon^2)$ for the hierarchical model specified in Equations~\eqref{eq:gls}--\eqref{eq:Delta},  Section~\ref{sec:fgls}, with a given matrix $\Phi$, we need to maximize the log-likelihood function given by the natural logarithm of the joint probability density function of the multivariate distribution $N(f_\beta(\mathbf{W}),\sigma_\varepsilon^2I+\Phi)$ evaluated at $y:=g(\hat C_U)\in\mathbb{R}^{m}$ with $m=|U|$, which is proportional to
\begin{eqnarray}\label{eq:fgls}
\ell(\beta,\sigma_\varepsilon^2)&=&-\log\det\left(\sigma_\varepsilon^2 I+\Phi\right)\\
  &&-(y- f_\beta(\mathbf{W}))^T(\sigma_{\varepsilon}^2 I + \Phi)^{-1}(y - f_\beta(\mathbf{W}))\nonumber
\end{eqnarray} 
By assuming a linear baseline regression function $f_\beta(\omega)=\omega^T\beta$ with $\beta\in\mathbb{R}^{q+1}$ and $\omega=(1,w)^T$, and by letting $\varphi = 2\log{\sigma_{\varepsilon}}\in\mathbb{R}$ to constrain $\sigma^{2}_{\varepsilon}=e^\varphi$ to be positive, the objective function in Equation~\eqref{eq:fgls} can be rewritten as
\begin{eqnarray}
        \ell(\beta,\varphi) &=& -\log{\det{(e^{\varphi}I + \Phi)}} \\
        &&- (y - \mathcal{W}\beta)^{T}(e^{\varphi}I + \Phi)^{-1}(y - \mathcal{W}\beta),~~~~~~\nonumber
    \end{eqnarray}
where $\mathcal{W}\in\mathbb{R}^{m\times (q+1)}$ contains 1s in the first columns and $\mathbf{W}$ in the other columns.  
We implement a gradient ascent algorithm (Algorithm~\ref{algo:optbaselinemeaserr}) to find the optimal values of $(\hat\beta,\hat\varphi)$ and thereby $(\hat\beta,\hat\sigma^2)$. Our gradient ascent algorithm uses an adaptive step size: if, on a step, the objective function decreases, we shrink the step size by a factor $a^{-1}$ and repeat the step; otherwise we increase the step size by a factor of $a$ ($a = 1$ would yield the standard non-adaptive gradient ascent). The algorithm stops if the relative difference of the values of the objective function in the current step and in the previous step is smaller than a threshold $s>0$, or if a maximum number of iterations $T$ is achieved.  In our computations, we use the OLS estimates of  $(\beta,\sigma^{2}_{\varepsilon})$ as start values. Moreover, we set an initial step size $b = 0.001$ with acceleration multiplier $a = 1.4$, and $s=10^{-7}$.  In practice, we observed that the algorithm typically converges within $40$ iterations, so we recommend $T\ge 40$.

\section{Proofs}\label{app:proofs}

\begin{proof}[\sc Proof of Theorem~\ref{theo:obscovconsist}]
The observed empirical covariance between variables $(i,j)$ in Equation~\eqref{eq:obscov} can be expanded as
\begin{eqnarray}
    \hat\Sigma_{ij} &=&   \frac{1}{n_{ij}}\sum_{r\in \mathcal{N}_{ij}}(X_i^{(r)}-m_i)(X_j^{(r)}-m_j)\\
    &=&  \frac{1}{n_{ij}}\sum_{r\in \mathcal{N}_{ij}}X_i^{(r)}X_j^{(r)} -m_j\frac{1}{n_{ij}}\sum_{r\in \mathcal{N}_{ij}}X_i^{(r)}~~~~\\
    &&-m_i\frac{1}{n_{ij}}\sum_{r\in \mathcal{N}_{ij}}X_j^{(r)}+m_im_j\nonumber
\end{eqnarray}

Since $\Sigma_{ii}<\infty$ for all $i$, we have $\E[(X_i^{(1)})^2]<\infty$ for all $i$, so by Cauchy-Schwarz Inequality we also have $\E[|X_i^{(1)}X_j^{(1)}|]\le\sqrt{\E[(X_i^{(1)})^2]\E[(X_j^{(1)})^2]}<\infty$ for all $(i,j)$. Thus, as $n_{ij}\to\infty$, we have $n_{ii},n_{jj}\to\infty$ because $n_{ij}\le n_{ii},n_{jj}$, and  by the Weak Law of Large Numbers,
\begin{eqnarray}
    \frac{1}{n_{ij}}\sum_{r\in\mathcal{N}_{ij}}X_i^{(r)}X_j^{(r)}~&\pto&~ \E[X_i^{(1)}X_j^{(1)}]~~~~\\
    \frac{1}{n_{ij}}\sum_{r\in\mathcal{N}_{ij}}X_j^{(r)}~&\pto&~ \E[X_i^{(1)}]\\
    \frac{1}{n_{ij}}\sum_{r\in\mathcal{N}_{ij}}X_j^{(r)}~&\pto&~ \E[X_j^{(1)}]\\
    \frac{1}{n_{ii}}\sum_{r\in\mathcal{N}_{ii}}X_i^{(r)} ~&\pto &~ \E[X_i^{(1)}]\\
    \frac{1}{n_{jj}}\sum_{r\in\mathcal{N}_{jj}}X_j^{(r)} ~&\pto &~ \E[X_j^{(1)}]
\end{eqnarray}
Therefore, regardless of whether $m_k:=\frac{1}{n_{kk}}\sum_{r\in \mathcal{N}_{kk}}X_k^{(r)}$ or $m_k:=\mathbb{E}[X_k^{(1)}]$ for all $k=1,\ldots,p$, by applying basic properties of sums and products of converging sequences, we obtain
\begin{eqnarray}
    \hat\Sigma_{ij} ~&\pto&~ \E[X_i^{(1)}X_j^{(1)}]-\E[X_i^{(1)}]\E[X_j^{(1)}]\\
    &&-\E[X_i^{(1)}]\E[X_j^{(1)}]+\E[X_i^{(1)}]\E[X_j^{(1)}]\nonumber\\
    &=& 
    \Sigma_{ij}
\end{eqnarray}
\end{proof}

\begin{proof}[\sc Proof of Lemma~\ref{lemma:covsigmaO}] By letting $Z_i^{(r)}=X_i^{(r)}-\mu_i$, where $\mu_i=\E[X_i^{(1)}]$, the observed empirical covariance between variables $(i,j)$ in Equation~\eqref{eq:obscov} with $m_i:=\mu_i$ reduces to
\begin{equation}
\hat\Sigma_{ij}=\frac{1}{n_{ij}} \sum_{r\in \mathcal{N}_{ij}} Z^{(r)}_{i}Z^{(r)}_{j}
\end{equation}

Thus,
\begin{eqnarray}
   &&\hspace{-5mm}\cov(\hat\Sigma_{ij},\hat\Sigma_{kl})\\ &=& \cov\left(\frac{1}{n_{ij}} \sum_{r\in \mathcal{N}_{ij}} Z^{(r)}_{i}Z^{(r)}_{j},~\frac{1}{n_{kl}} \sum_{t\in \mathcal{N}_{kl}} Z^{(t)}_{k}Z^{(t)}_{l}\right)~~~~~~~~~~~\\
    &=& \frac{1}{n_{ij}}\frac{1}{n_{kl}}\cov\left(\sum_{r\in \mathcal{N}_{ij}} Z^{(r)}_{i}Z^{(r)}_{j},~ \sum_{t\in \mathcal{N}_{kl}} Z^{(t)}_{k}Z^{(t)}_{l}\right)\\
    &=& \frac{1}{n_{ij}}\frac{1}{n_{kl}}\sum_{r\in \mathcal{N}_{ij}}\sum_{t\in \mathcal{N}_{kl}}\cov\left( Z^{(r)}_{i}Z^{(r)}_{j},~  Z^{(t)}_{k}Z^{(t)}_{l}\right)\\
    &=& \frac{1}{n_{ij}}\frac{1}{n_{kl}}\sum_{r\in \mathcal{N}_{ij}\cap \mathcal{N}_{kl}} \cov\left( Z^{(r)}_{i}Z^{(r)}_{j},~  Z^{(r)}_{k}Z^{(r)}_{l}\right)\label{eq:cov0}\\ 
    &=& \frac{n_{ijkl}}{n_{ij} n_{kl}}   \cov\left( Z^{(1)}_{i}Z^{(1)}_{j},~  Z^{(1)}_{k}Z^{(1)}_{l}\right)\label{eq:covn},
\end{eqnarray}
where Equation~\eqref{eq:cov0} is due to independence which implies $\cov\left( Z^{(r)}_{i}Z^{(r)}_{j},~  Z^{(t)}_{k}Z^{(t)}_{l}\right)=0$ if $r\neq t$, and Equation~\eqref{eq:covn} is due to identical distributions. 
\end{proof}

~

\begin{proof}[\sc Proof of Theorem~\ref{theo:asynormincomplete}]  
Let $\hat{\mu}$ and $\hat{\Sigma}_O$ be the observed empirical mean vector and covariance matrix, where
\begin{equation}
    \hat\mu_i=\frac{1}{n_{ii}}\sum_{r\in \mathcal{N}_{ii}} X^{(r)}_i  
\end{equation}
\begin{equation}
    \hat\Sigma_{ij}=\frac{1}{n_{ij}}\sum_{r\in \mathcal{N}_{ij}}(X_i^{(r)}-\hat\mu_i)(X_j^{(r)}-\hat\mu_j)
\end{equation}

Moreover, define
\begin{equation}\label{eq:sigtild}
    \tilde\Sigma_{ij}=\frac{1}{n_{ij}}\sum_{r\in \mathcal{N}_{ij}}(X_i^{(r)}-\mu_i)(X_j^{(r)}-\mu_j)
\end{equation}
For $t=1,...,K$, define the $p\times p$ matrix  $\tilde\Sigma^{(t)}$ with $V_t\times V_t$ portion equal to
\begin{equation}
\tilde\Sigma^{(t)}_{V_tV_t}=\frac{1}{n_t}\sum_{r\in R_t} (X^{(r)}_{V_t}-\mu_{V_t})(X^{(r)}_{V_t}-\mu_{V_t})^T,
\end{equation}
where $R_t=\{r:V^{(r)}=V_t\}$ and $n_t=|R_t|$, while all other entries of $\tilde\Sigma^{(t)}$ are zero. Since $\E[(X_i^{(1)})^4]<\infty$ for all $i$, by Cauchy-Schwarz Inequality we have $\E[(X_i^{(1)}X_j^{(1)})^2]\le\sqrt{\E[(X_i^{(1)})^4]\E[(X_j^{(1)})^4]} < \infty, \forall (i,j)$, so by the Central Limit Theorem, as $n_t\to\infty$, 
\begin{equation}\label{eq:conv1}
\sqrt{n_t}\left(\tilde\Sigma^{(t)}_{\bar U}-\Sigma_{\bar U}^*\right)\dto N(0,H^{(t)})
\end{equation}
where $\Sigma_{\bar U\cap (V_t\times V_t)}^*=\Sigma_{\bar U\cap (V_t\times V_t)}$, $\Sigma^*_{\bar U\cap (V_t\times V_t)^C}=0$, and by letting $Z^{(1)}_i:=X^{(1)}_i-\mu_i$ we obtain
\begin{equation}
H^{(t)}_{(i,j),(k,l)} = \cov\big(Z^{(1)}_iZ^{(1)}_j,Z^{(1)}_kZ^{(1)}_l\big) 
\end{equation}
if $(i,j),(k,l)\in \bar U\cap (V_t\times V_t)$, and $H^{(t)}_{(i,j),(k,l)}=0$ otherwise.  
By letting $I^{(t)}_{ij}:=\mathcal{I}(\{i, j\}\subseteq V_t)$, where $\mathcal{I}()$ is the indicator function, for $(i,j)\in \bar U$ Equation~\eqref{eq:sigtild} can be rewritten as
\begin{eqnarray}
\tilde\Sigma_{ij} 
&=&  \sum_{t=1}^K \frac{I^{(t)}_{ij}\cdot n_t}{\sum_{h=1}^K I^{(h)}_{ij}\cdot n_h} \tilde\Sigma^{(t)}_{ij} 
\end{eqnarray} 
Note that, $n_t=\lceil \pi_t\cdot n\rfloor=\pi_t\cdot n+\xi_{t,n}$, where $\pi_1,...,\pi_K$ are fixed constants such that $\sum_{t=1}^K\pi_t=1$, and $\xi_{t,n}\in (-0.5,0.5)$ for all $t,n$, for $(i,j)\in \bar U$, so
\begin{eqnarray}
&&\hspace{-8mm}\sqrt{n}\left(\tilde\Sigma_{ij}-\Sigma_{ij}\right)  \\ 
&=&   \sqrt{n}\sum_{t=1}^K \frac{I^{(t)}_{ij}\cdot n_t}{\sum_{h=1}^K I^{(h)}_{ij}\cdot n_h}  \left(\tilde\Sigma^{(t)}_{ij}-\Sigma_{ij}\right)\nonumber\\
&=&   \sum_{t=1}^K \frac{I^{(t)}_{ij}\cdot \sqrt{n n_t}}{\sum_{h=1}^K I^{(h)}_{ij}\cdot n_h}  \sqrt{n_t}\left(\tilde\Sigma^{(t)}_{ij}-\Sigma_{ij}\right)\\
&=&  \sum_{t=1}^K Q_{ij}^{(t)}(n) \sqrt{n_t}\left(\tilde\Sigma^{(t)}_{ij}-\Sigma_{ij}\right),
\end{eqnarray}
where
\begin{equation}
Q_{ij}^{(t)}(n):=\frac{I^{(t)}_{ij}\cdot \sqrt{ \pi_t+\xi_{t,n}/n}}{\sum_{h=1}^K I^{(h)}_{ij}\cdot (\pi_h+\xi_{h,n}/n)} 
\end{equation}
and 
\begin{equation}\label{eq:conv2}
\limn Q_{ij}^{(t)}(n) = Q_{ij}^{(t)}:=\frac{I^{(t)}_{ij}\sqrt{\pi_t}}{\sum_{h=1}^KI^{(h)}_{ij}\pi_h}
\end{equation}
Therefore, by combining Equations~\eqref{eq:conv1}--\eqref{eq:conv2} via Slutsky's Theorem and properties of sums of converging independent sequences of Gaussian random vectors, we obtain
\begin{eqnarray}
\sqrt{n}\left(\tilde\Sigma_{\bar U}-\Sigma_{\bar U}\right)&=& \sum_{t=1}^K Q_{\bar U}^{(t)}(n)\odot\sqrt{n_t}\left(\tilde\Sigma^{(t)}_{\bar U} -\Sigma^*_{\bar U}\right)~~~~~\nonumber\\
&\dto & N\left(0,H\right)\label{eq:asnormtilde}
\end{eqnarray}
where $\odot$ denotes the entrywise product, and
\begin{equation}
H~=~\sum_{t=1}^K {\rm diag}\left(Q^{(t)}_{\bar U}\right) H^{(t)}{\rm diag}\left(Q^{(t)}_{\bar U}\right)
\end{equation}
Thus, for $(i,j),(k,l)\in \bar U$, we have
\begin{eqnarray}
H_{(i,j),(k,l)} &=& \sum_{t=1}^K Q^{(t)}_{ij}Q^{(t)}_{kl} H^{(t)}_{(i,j),(k,l)}\\
&=& c_{ijkl} \cov\big(Z^{(1)}_i Z^{(1)}_j, Z^{(1)}_k Z^{(1)}_l\big),~~~~~~
\end{eqnarray}
where 
\begin{eqnarray}
c_{ijkl}&:=&\frac{\sum_{t=1}^KI^{(t)}_{ij}I^{(t)}_{kl}\pi_t}{\left(\sum_{t=1}^KI^{(t)}_{ij}\pi_t\right)\left(\sum_{t=1}^KI^{(t)}_{kl}\pi_t\right)}
\end{eqnarray}
Now, note that  
\begin{eqnarray}\label{eq:convX1}
\hat\Sigma_{ij}-\tilde\Sigma_{ij} 
&=& (\hat\mu_i-\mu_i)(\hat\mu_j-\mu_j)\\
&&-(\breve\mu_i-\mu_i)[(\hat\mu_j-\mu_j)+(\breve\mu_j-\mu_j)]\nonumber
\end{eqnarray}
where $\breve\mu_i=\frac{1}{n_{ij}}\sum_{r\in \mathcal{N}_{ij}}X_i^{(r)}$ and $\breve\mu_j=\frac{1}{n_{ij}}\sum_{r\in \mathcal{N}_{ij}}X_j^{(r)}$.  
Since $\Sigma_{ii},\Sigma_{jj}<\infty$, by the Central Limit Theorem, as $n\to\infty$,  
\begin{eqnarray}
\sqrt{n_{ii} }(\hat\mu_i-\mu_i)&\dto& N(0,\Sigma_{ii})\\
\sqrt{n_{ij}}(\breve\mu_i-\mu_i)&\dto& N(0,\Sigma_{ii})
\end{eqnarray}
and by the Weak Law of Large Numbers, 
\begin{eqnarray}
\hat\mu_j-\mu_j&\pto& 0\\
\breve\mu_j-\mu_j&\pto& 0
\end{eqnarray}
for all $(i,j)\in \bar U$. Moreover, for all $(i,j)\in \bar U$, 
\begin{eqnarray}
\frac{n}{n_{ij}}&=&\frac{n}{\sum_{t=1}^K I^{(t)}_{ij}\cdot n_t}\\
&=&\frac{n}{\sum_{t=1}^K I^{(t)}_{ij}\cdot (\pi_t\cdot n+\xi_{t,n})}\\
&=&\frac{1}{\sum_{t=1}^K I^{(t)}_{ij}\cdot (\pi_t+\xi_{t,n}/n)}\\
&\to& \frac{1}{\sum_{t=1}^K I^{(t)}_{ij}\cdot \pi_t} <\infty\label{eq:convX2}
\end{eqnarray}
as $n\to\infty$. Thus, by combining Equations~\eqref{eq:convX1}--\eqref{eq:convX2} via Slutsky's Theorem, we obtain that for all $(i,j)\in\bar U$, as $n\to\infty$,
\begin{eqnarray}
&&\hspace{-10mm}\sqrt{n}(\hat\Sigma_{ij}-\tilde\Sigma_{ij})=\sqrt{n}(\hat\mu_i-\mu_i)(\hat\mu_j-\mu_j)\\
&&~~~~-\sqrt{n}(\breve\mu_i-\mu_i)[(\hat\mu_j-\mu_j)+(\breve\mu_j-\mu_j)]\nonumber\\
&=&\sqrt{n_{ii}}(\hat\mu_i-\mu_i)\sqrt{\frac{n}{n_{ii}}}(\hat\mu_j-\mu_j)\\
&&-\sqrt{n_{ij}}(\breve\mu_i-\mu_i)\sqrt{\frac{n}{n_{ij}}}[(\hat\mu_j-\mu_j)+(\breve\mu_j-\mu_j)]~~~\nonumber\\
&\pto & 0
\end{eqnarray}
that is, as $n\to\infty$,
\begin{equation}\label{eqn: Sigma hat minus Sigma tilde U}
    \sqrt{n} (\hat{\Sigma}_{\bar U} - \tilde{\Sigma}_{\bar U}) \pto
    0
\end{equation}
By Slutsky's Theorem, Equations~\eqref{eq:asnormtilde} and \eqref{eqn: Sigma hat minus Sigma tilde U} imply
\begin{eqnarray}
    \sqrt{n}(\hat\Sigma_{\bar U}-\Sigma_{\bar U})\hspace{-2mm} &=& \hspace{-2mm}\sqrt{n} (\hat{\Sigma}_{\bar U} - \tilde\Sigma_{\bar U}) + \sqrt{n} (\tilde{\Sigma}_{\bar U} - \Sigma_{\bar U}) ~~~~~~~~~~~\\
    \hspace{-2mm}&\dto & \hspace{-2mm}N(0,H)\nonumber
\end{eqnarray}
By applying the Multivariate Delta Method
\begin{equation}
\sqrt{n}(\hat{C}_U - C_U)\dto  N(0,JHJ^{T}) 
\end{equation}
where $J$ is the $|U|\times |\bar U| $ Jacobian matrix with entry
\begin{eqnarray}
\hspace{-8mm}&&\hspace{-8mm}J_{(i,j),(l,k)} = \frac{\partial C_{ij}}{\partial\Sigma_{lk}}=\frac{\partial (\Sigma_{ij}/\sqrt{\Sigma_{ii}\Sigma_{jj}})}{\partial\Sigma_{lk}} \\
&=& \hspace{-2mm}{\small \left\{ \begin{array}{cc}
 \hspace{-2mm}\frac{1}{\sqrt{\Sigma_{ii}\Sigma_{jj}}}    &\hspace{-2mm} {\rm if}~ l=i,k=j, i\neq j  \:or \:k=i,l=j, i\neq j\\
\hspace{-2mm} 0   &\hspace{-2mm} {\rm if}~i= j ~\text{or}~ (l,k)\notin\{(i,j),(i,i),(j,j)\}\\
\hspace{-2mm} -\frac{\Sigma_{ij}}{2\Sigma_{ii}^{3/2}\sqrt{\Sigma_{jj}}}    & \hspace{-2mm} {\rm if}~l=k=i, ~i\neq j \\
\hspace{-2mm} -\frac{\Sigma_{ij}}{2\Sigma_{jj}^{3/2}\sqrt{\Sigma_{ii}}}    & \hspace{-2mm}{\rm if}~l=k=j, ~i\neq j \\ 
\end{array}\right.\nonumber}
\end{eqnarray}
for $(i,j)\in U$ and $(k,l)\in \bar U$. 
Finally, by applying one more time the Multivariate Delta Method, we obtain 
\begin{equation}
    \sqrt{n} (g(\hat{C}_U) - g(C_U)) \dto N(0,FJHJ^{T}F^{T})
\end{equation}
where $F$ is the $|U|\times |U|$ Jacobian diagonal matrix of the Fisher transformation, with diagonal entry \begin{equation}
F_{(i,j),(i,j)} = g'(C_{ij}) =  (1-C_{ij}^2)^{-1}
\end{equation} for $(i,j)\in U$. The assumption of $C$ to be positive definite ensures $|C_{ij}|<1$, and thereby $0<|F_{(i,j),(i,j)}|<\infty$ for all $(i,j)\in U$.
\end{proof}

\begin{proof}[\sc{Proof of Corollary~\ref{coro:psihat}}]~
\begin{enumerate}[(i).]
\item By Theorem~\ref{theo:asynormincomplete} and the Continuous Mapping Theorem, we have \begin{equation}
\hspace{-1.5mm}\sqrt{n}\Psi^{-1/2}(g(\hat C_U)-g(C_U))\dto Z\sim N(0,I_{|U|})
\end{equation}
where $I_{|U|}$ is the $|U|\times|U|$ identity matrix. If $\hat\Psi\pto \Psi$, by the Continuous Mapping Theorem we obtain $\hat\Psi^{-1/2}\Psi^{1/2}\pto I_{|U|}$. Therefore, by Slutsky's Theorem,
\begin{eqnarray}
&&\hspace{-10mm}\sqrt{n}\hat\Psi^{-1/2}(g(\hat C_U)-g(C_U))\\
&=& \hat\Psi^{-1/2}\Psi^{1/2}  \sqrt{n}\Psi^{-1/2}(g(\hat C_U)-g(C_U))~~~\nonumber\\
&\dto & I_{|U|} Z ~\sim N(0,I_{|U|})
\end{eqnarray}

\item 
Without loss of generality, consider the $(1,2),(3,4)$ entry of $H$ (Equation~\eqref{eq:Hemp})
\begin{eqnarray}
\hat H_{(1,2),(3,4)} &=& \hat c_{1234}(M_{1234}-M_{12}M_{34})~~~~~
\end{eqnarray}
and assume $\hat c_{1234}>0$. First, note that, using the notation defined in the proof of Theorem~\ref{theo:asynormincomplete},
\begin{eqnarray}
    \frac{n_{ijkl}}{n} &=& \frac{\sum_{t=1}^K I^{(t)}_{ij} I^{(t)}_{kl} n_t}{n}\\
    &=&\frac{\sum_{t=1}^K I^{(t)}_{ij} I^{(t)}_{kl} (\pi_t\cdot n+\xi_{t,n})}{n}~~\\
&=&\sum_{t=1}^K I^{(t)}_{ij} I^{(t)}_{kl} (\pi_t+\xi_{t,n}/n)\\
&\to& \sum_{t=1}^K I^{(t)}_{ij} I^{(t)}_{kl} \pi_t
\end{eqnarray}
as $n\to\infty$, and combining this limit with the one in Equation~\eqref{eq:convX2}, we obtain
\begin{eqnarray}
\hat c_{ijkl} &=&\frac{n}{n_{ij}}\frac{n}{n_{kl}}\frac{n_{ijkl}}{n}\\
&\to& \frac{\sum_{t=1}^KI^{(t)}_{ij}I^{(t)}_{kl}\pi_t}{\left(\sum_{t=1}^KI^{(t)}_{ij}\pi_t\right)\left(\sum_{t=1}^KI^{(t)}_{kl}\pi_t\right)}~~~~~\\
&=&c_{ijkl}
\end{eqnarray}
as $n\to\infty$. Next, by applying the mathematical identity
\begin{equation}
\prod_{k=1}^N(a_k+b_k)~=~\sum_{A\subseteq \{1,...,N\}}\prod_{i\in A}a_i\prod_{j\notin A}b_j
\end{equation}
with the convention $\prod_{i\in\emptyset}a_i=1$, we obtain
\begin{eqnarray}
&&\hspace{-15mm} M_{1234} = 
\frac{1}{n_{1234}}\sum_{r\in\mathcal{N}_{1234}} \prod_{k=1}^4(X_{k}^{(r)}-M_k)\\
&&\hspace{-15mm} =\frac{1}{n_{1234}}\sum_{r\in\mathcal{N}_{1234}}\sum_{A\subseteq\{1,2,3,4\}} \prod_{i\in A} X_{i}^{(r)} \prod_{j\notin A} (-M_j)~~~\\
&&\hspace{-15mm} = \sum_{A\subseteq {\{1,2,3,4\}}} \prod_{j\notin A} (-M_j)\frac{1}{n_{1234}} \sum_{r\in\mathcal{N}_{1234}} \prod_{i\in A} X_{i}^{(r)}
\end{eqnarray}
where $\mathcal{N}_{1234}=\mathcal{N}_{12}\cap\mathcal{N}_{34}$ and $n_{1234}=|\mathcal{N}_{1234}|$. Under the assumptions of Theorem~\ref{theo:asynormincomplete}, we have  $\E[(X_i^{(1)})^2]<\infty$ and $\E[(X_i^{(1)}X_j^{(1)})^2]<\infty$, $\forall (i,j)$, so by Cauchy-Schwarz Inequality, 
\begin{eqnarray}
&&\hspace{-7mm}\left(\E\left[\left\vert\prod_{i\in A} X_{i}^{(1)}\right\vert \right]\right)^2\\
&&\hspace{-7mm}\le \E\left[\prod_{i\in A\cap\{1,2\}} (X_{i}^{(1)})^2\right]\E\left[\prod_{j\in A\cap\{3,4\}} (X_{j}^{(1)})^2\right]<\infty\nonumber
\end{eqnarray}
for all $A\subseteq \{1,2,3,4\}$. 
Thus, by the Weak Law of Large Numbers, as $n\to\infty$,
\begin{eqnarray}
M_j &\pto& \mu_j\\
\frac{1}{n_{1234}} \sum_{r\in\mathcal{N}_{1234}} \prod_{i\in A} X_{i}^{(r)}&\pto& \E\left[\prod_{i\in A} X_{i}^{(1)}\right]~~~~~~~~~~~
\end{eqnarray}
Thus, by basic properties of sums and products of converging sequences we obtain
\begin{eqnarray}
    M_{1234} 
    \hspace{-2mm} &\pto& \hspace{-3mm} \sum_{A\subseteq {\{1,2,3,4\}}} \prod_{j\notin A} (-\mu_j)\E\left[\prod_{i\in A} X_{i}^{(1)}\right]~~~~~~\\
&=& \E\left[\prod_{k=1}^4(X_k^{(1)}-\mu_k)\right]
\end{eqnarray}
Similarly,
\begin{eqnarray}
    \hspace{-12mm}M_{12} 
    \hspace{-2mm}&=&\hspace{-2mm} \frac{1}{n_{12}}\sum_{r\in\mathcal{N}_{12}} (X_1^{(r)}-M_1)(X_2^{(r)}-M_2) ~~\\  
   \hspace{-2mm} &\pto & \hspace{-2mm}
    \E[(X_1^{(1)}-\mu_1)(X_2^{(1)}-\mu_2)]
\end{eqnarray}
Therefore,
\begin{equation}
    \hat H_{(1,2),(3,4)}\pto H_{(1,2),(3,4)}
\end{equation}
Similarly, we can prove $\hat H_{(i,j),(k,l)}\pto H_{(i,j),(k,l)}$, for all $(i,j),(k,l)\in\bar U$. 

Since the entries of the matrices $J$ and $F$ are continuous functions of $\Sigma$, and each entry of the observed sample covariance matrix $\hat{\Sigma}_O$ converges in probability to the corresponding entry of $\Sigma_O$ (Theorem~\ref{theo:obscovconsist}), by the Continuous Mapping Theorem, $\hat{J} \overset{P}{\rightarrow} J$ and $\hat{F} \overset{P}{\rightarrow} F$. Finally, by properties of sums and products of converging sequences,
    \begin{equation}        \hat\Psi=\hat{F}\hat{J}\hat{H}\hat{J}^{T}\hat{F}^{T} \overset{P}{\rightarrow}
        FJHJ^{T}F^{T} = \Psi
    \end{equation}

\item Let $Z := X^{(r)} - \mu\sim N(0,\Sigma)$. Then,  Equation~\eqref{eq:Hentry} can be written as
\begin{eqnarray}
   && \hspace{-5mm} H_{(i,j),(k,l)} = c_{ijkl} \cov\big(Z_{i}Z_{j}, ~ Z_{k}Z_{l}\big)\\
    &=& c_{ijkl} \left(\E[Z_{i}Z_{j}Z_{k}Z_{l}] - \E[Z_{i}Z_{j}]\E[Z_{k}Z_{l}]\right)~~~~~~~~~~~\\
    &=& c_{ijkl} \left(\E[Z_{i}Z_{j}Z_{k}Z_{l}] - \Sigma_{ij}\Sigma_{kl}\right)
\end{eqnarray}
By Isserlis' Theorem,
\begin{eqnarray}    \E[Z_{i}Z_{j}Z_{k}Z_{l}] \hspace{-2mm}&=&\hspace{-2mm} \E[Z_{i}Z_{j}]\E[Z_{k}Z_{l}] \\&&\hspace{-2mm}+ \E[Z_{i}Z_{k}]\E[Z_{j}Z_{l}]\nonumber \\&&\hspace{-2mm}+ \E[Z_{i}Z_{l}]\E[Z_{j}Z_{k}]\nonumber\\
    \hspace{-2mm}&=&\hspace{-2mm} \Sigma_{ij}\Sigma_{kl}+\Sigma_{ik}\Sigma_{jl}+\Sigma_{il}\Sigma_{jk}~~~~~~~~
\end{eqnarray}
so
\begin{eqnarray}
    H_{(i,j),(k,l)} &=& c_{ijkl}  \left(\Sigma_{ik}\Sigma_{jl}+\Sigma_{il}\Sigma_{jk}\right),~~~~
\end{eqnarray}
and since, by Theorem~\ref{theo:obscovconsist},  $\hat\Sigma_{ij}\pto\Sigma_{ij}$ for all $(i,j)\in\bar U$,  by basic properties of sums and products of converging sequences we obtain 
\[
\breve H_{(i,j),(k,l)}\pto H_{(i,j),(k,l)}
\]
for all $(i,j),(k,l)\in\bar U$. 
From part (ii), we have $\hat J\pto J$ and $\hat F\pto F$, so by the properties of sums and products of converging sequences we finally obtain
    \begin{equation}        \tilde\Psi=\hat{F}\hat{J}\breve H\hat{J}^{T}\hat{F}^{T} \overset{P}{\rightarrow}
        FJHJ^{T}F^{T} = \Psi
    \end{equation}
\end{enumerate}
\end{proof}

\section{Additional simulations}\label{app:addsims}
\subsection{Verifying Theorem~\ref{theo:asynormincomplete} and Corollary~\ref{coro:psihat}}\label{app:veriftheo}
\begin{figure}[t!]
    \centering
    \includegraphics[width=1\columnwidth]{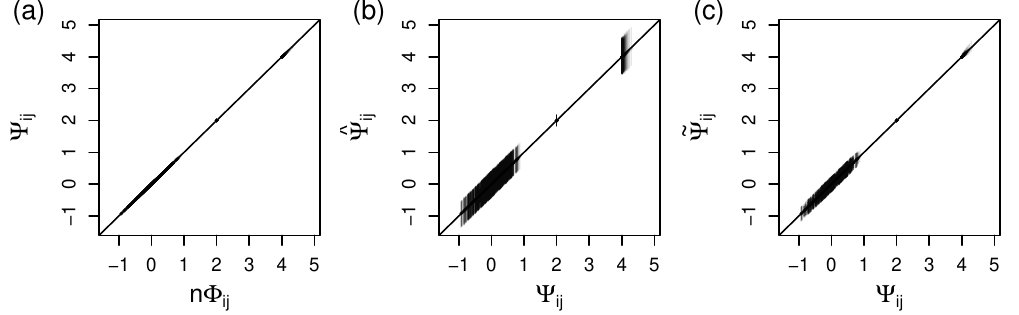}\vspace{-3mm}
    \caption{{\bf (a)} Entries of the asymptotic covariance matrix $\Psi$ versus the corresponding entries of the nonasymptotic covariances of Fisher-transformed observed sample correlations. {\bf (b)} Tolerance intervals (95\%) of the entries of the empirical estimator $\hat\Psi$ (Equation~\eqref{eq:psiemphat}) against the corresponding entries of $\Psi$.  {\bf (c)} Tolerance intervals (95\%) of the entries of the (Gaussian) estimator $\tilde\Psi$ (Equation~\eqref{eq:psigausshat}) against the corresponding entries of $\Psi$.
    }
    \label{fig:asympsim}
\end{figure}

Let $\hat C_U$ be the vector of observed empirical correlations computed from partially observed $p$-dimensional random vectors $X_1,...,X_n\iid N(0,\Sigma)$, and let $g(\omega)$ be the Fisher transformation (Equation~\eqref{eq:fishertransf}). Let $\Phi=\cov(g(\hat C_U))$, which we compute via Monte Carlo integration ($10^6$ draws). According to Theorem~\ref{theo:asynormincomplete}, we have $\Phi\approx \Psi/n$.  In Figure~\ref{fig:asympsim}, we show the results of a simulation implemented in the same way as in Section~\ref{sec:simGround} with $p=40$,  $n=5000$, $\eta=0.45$, $K=4$. In Figure~\ref{fig:asympsim}(a), we can see that $\Psi$ approximates $n\Phi$ well, while in Figures~\ref{fig:asympsim}(b) and ~\ref{fig:asympsim}(c) we show that the empirical estimator $\hat\Psi$  (Equation~\eqref{eq:psiemphat}) and the one assuming Gaussian data $\tilde\Psi$ (Equation~\eqref{eq:psigausshat}) given in Corollary~\ref{coro:psihat} both reasonably estimate $\Psi$, although $\tilde\Psi$ appears to have smaller variance.

\subsection{Accuracy of tuning parameter selection in \glsname{}}\label{app:simfgls}
\begin{figure}[t!]
    \centering
    \includegraphics[width=1\columnwidth]{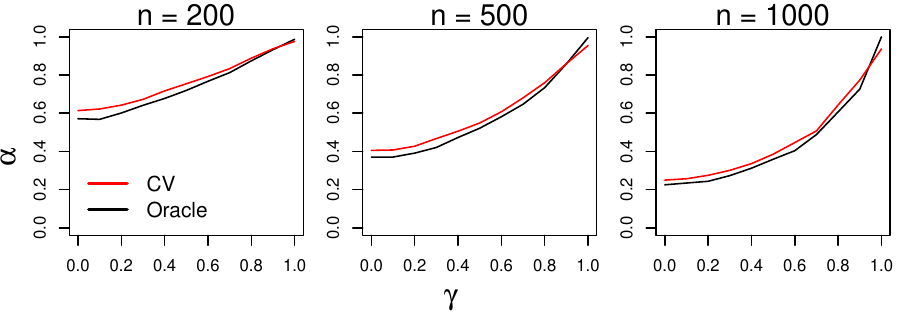}\vspace{-3mm}
    \caption{{\bf Tuning parameter selection in \glsname{} via 10-CV}. 
    Average tuning parameter $\alpha_{\rm cv}$ selected in \glsname{} via $10$-CV (red) and oracle counterpart $\alpha_{\rm or}$ (black) are plotted against $\gamma$ (Equation~\eqref{eq:gamma}), for various scenarios with number of data sets $K=2$, missingness proportion $\eta=0.45$, number of variables $p=50$, and total sample size $n=200,500,1000$. The average $\alpha_{\rm cv}$ and $\alpha_{\rm or}$ are very close, increase with $\gamma$, and decrease with $n$.
}
    \label{fig:CV v Oracle FGLS}
\end{figure}

We present the results of a simulation analogous to the one presented in Section~\ref{sec:simCV}, but using \glsname{} (Equation~\eqref{eq:fgls}) for $p=50$,  $n=200,500,1000$, $K=2$, and $\eta=0.45$. In Figure~\ref{fig:CV v Oracle FGLS}, we can see that the average $\alpha_{\rm cv}$ and $\alpha_{\rm or}$ are very close, and they both increase with $\gamma$ and decrease with $n$.

\subsection{Additional methods comparisons}\label{app:addsimmodelcomp}
\begin{figure}[t!]
    \label{fig:simcomparemethods}
    \includegraphics[width=1\columnwidth]{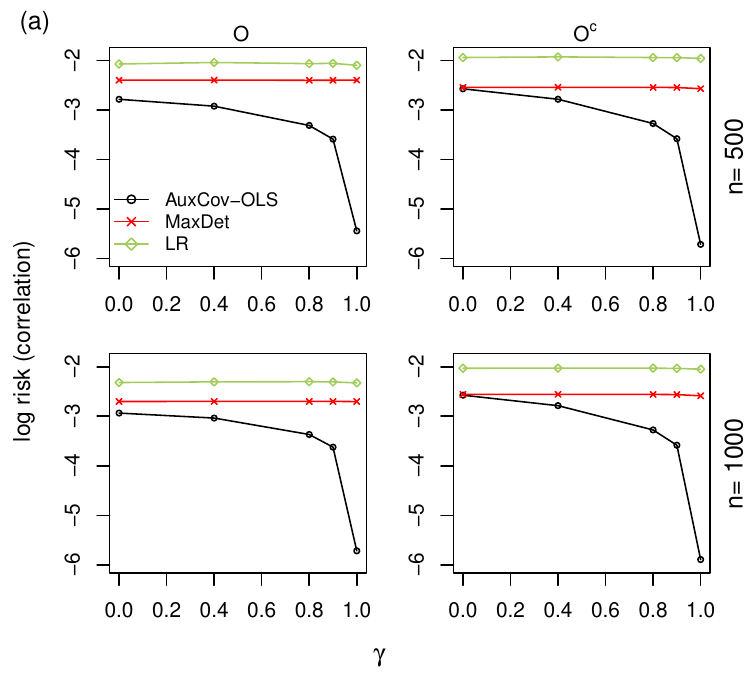}\\
    \includegraphics[width=1\columnwidth]{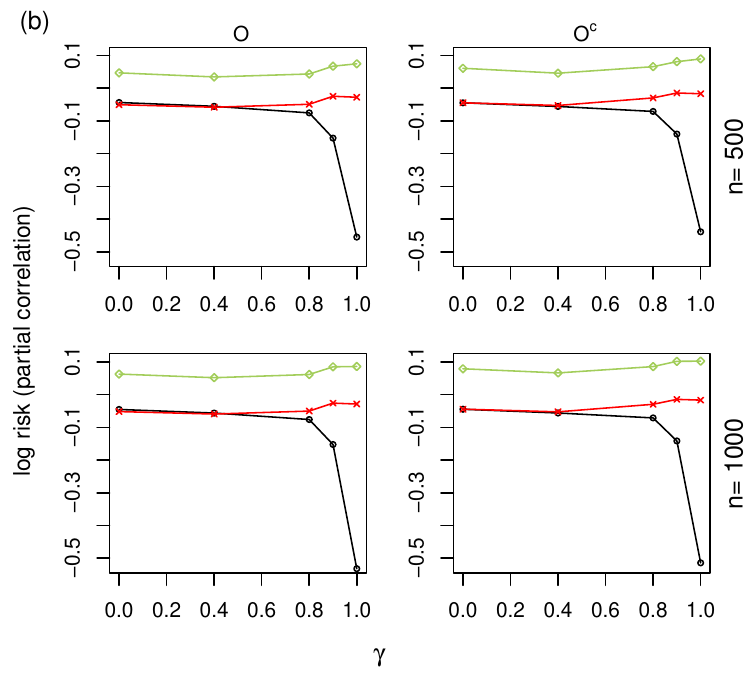
    }\vspace{-5mm}    \caption{{\bf Methods comparison.}
    Performance of \olsname{}, MaxDet, and LR at recovering {\bf (a)} correlations and  {\bf (b)} 
 partial correlations for $\gamma\in[0,1]$, $p=100$, $n=500,1000$, $K=2$, $\eta=0.45$. \olsname{} outperforms MaxDet and LR in most conditions, especially for larger values of $\gamma$.}
\label{fig:simcomparemethodsApp}
\end{figure}

We repeated the simulation presented in Section~\ref{sec:simCompare} but with a larger number of variables, $p=100$, and a larger proportion of missingness, $\eta=0.45$, using only \olsname{} (the computationally fastest \methname{} variant), MAD, and LR. Results are shown in Figure~\ref{fig:simcomparemethodsApp} and are analogous to Figure~\ref{fig:comparemethodssim}.

\bibliographystyle{unsrt}
\bibliography{references}

\end{document}